\documentclass[english,twoside,a4paper,10pt]{article}

\usepackage[latin1]{inputenc}
\usepackage[T1]{fontenc}
\usepackage{amsmath}
\usepackage{amsfonts}
\usepackage{graphicx}
\usepackage[numbers,sort&compress]{natbib}
\usepackage{subfigure}
\usepackage{a4wide}
\usepackage{amssymb}
\usepackage{fancyhdr}
\usepackage{mathrsfs}
\usepackage[toc,page]{appendix}

\begin{document}

\title{\bf Mimetic gravity: a review of recent developments and applications to cosmology and astrophysics}

\author{
L. Sebastiani\footnote{E-mail address: lorenzo.sebastiani@unitn.it}\,\,\,
\\ \\
\begin{small}
Department of General \& Theoretical Physics and Eurasian Center for Theoretical Physics,
\end{small} \\
\begin{small} 
Eurasian National University, 010008 Astana, Satpayev Str. 2, Kazakhstan
\end{small}\\
\\ \\
S. Vagnozzi\footnote{E-mail address: sunny.vagnozzi@fysik.su.se}\,\,\,
\\ \\
\begin{small}
Niels Bohr Institute, University of Copenhagen, Blegdamsvej 17, 2100 K\o benhavn \O, Denmark
\end{small}\\
\\ \\
R. Myrzakulov\footnote{E-mail address: rmyrzakulov@gmail.com}\,\,\,
\\
\begin{small}
Department of General \& Theoretical Physics and Eurasian Center for Theoretical Physics,
\end{small} \\
\begin{small} 
Eurasian National University, 010008 Astana, Satpayev Str. 2, Kazakhstan
\end{small}\\
}

\date{}

\maketitle

\begin{abstract}
Mimetic gravity is a Weyl-symmetric extension of General Relativity, related to the latter by a singular disformal transformation, wherein the appearance of a dust-like perfect fluid can mimic cold dark matter at a cosmological level. Within this framework, it is possible to provide an unified geometrical explanation for dark matter, the late-time acceleration, and inflation, making it a very attractive theory. In this review, we summarize the main aspects of mimetic gravity, as well as extensions of the minimal formulation of the model. We devote particular focus to the reconstruction technique, which allows the realization of any desired expansionary history of the Universe by an accurate choice of potential, or other functions defined within the theory (as in the case of mimetic $f(R)$ gravity). We briefly discuss cosmological perturbation theory within mimetic gravity. As a case study within which we apply the concepts previously discussed, we study a mimetic Ho\v{r}ava-like theory, of which we explore solutions and cosmological perturbations in detail. Finally, we conclude the review by discussing static spherically symmetric solutions within mimetic gravity, and apply our findings to the problem of galactic rotation curves. Our review provides an introduction to mimetic gravity, as well as a concise but self-contained summary of recent findings, progresses, open questions, and outlooks on future research directions.
\end{abstract}

\tableofcontents

\newpage

\section{Introduction}

The past decade has seen the astounding confirmation of the ``dark Universe'' picture, wherein the energy budget of our Universe is dominated by two dark components: dark matter and dark energy \cite{du1,du2,du3,du4,du5,du6,du7,du8,du9,du10,du11,du12,du13,du14,du15,du16,du17,du18,du19,du20,
du21,du22,du23,du24,du25,du26,du27,du28,du29,du30,du31,du32,du33,du34,du35,du36,du37,du38,du39,
du40,du41,du42,du43,du44,du45,du46,du47,du48,du49,du50}. The race to determine the nature and origin of these components is in progress both on the observational and theoretical fronts. Theories of modified gravity appear quite promising in this respect, particularly given that gravity remains the least understood of the four fundamental forces. For an incomplete list of comprehensive reviews, as well as seminal works on the subject, we refer the reader to \cite{curvature,turner,vollick,mtheory,kawa,chiba,ln,easson,cardone,nojiriodintsov1,nojiriodintsov2,
woodard,amendola,troisi,nojiriodintsov3,clifton,delacapozziello,bambacapozziello} and references therein. \\

A particularly interesting theory of modified gravity which has emerged in the past few years is mimetic gravity \cite{m1}. In mimetic gravity, as well as minimal modifications thereof, it is possible to describe the dark components of the Universe as a purely geometrical effect, without the need of introducing additional matter fields. In the past three years, interest in this theory has grown rapidly, with over 90 papers following up on the original idea or at least touching on it in some way \cite{golovnev,barvinsky,sheykin,m2,momeni,chaichian,malaeb,harko,battefeld,deruelle,altaibayeva,nompla,chamseddine,
saadi,capela,mirzagholi,leon,haghani,matsumoto,gudekli,yuan,sebastiani,jacobson,speranza,astashenok,myrzakulov,domenech,
zerbini,bufalo,silva,moraes,raza,deffayet,khalifeh,arroja,ramazanov,kunz,mukohyama,shiravand,kairat,guendelman1,kan,
oikonomou,guendelman2,rabochaya,oikonomousingular,myrzakul,nurgissa,asenjo,sebastianicurves,oikonomouaccelerating,
guendelman3,oikonomouanti,sebastianiscalar,guendelman4,koshelev,astashenokneutron,husain,bartolo,hammer,oikonomouapss,
cognola,higherorder10,sebastianinonlocal,chamseddineinhomogeneous,matarrese,odintsovunimodular,babichev,
odintsovum,saridakis,langlois,weiner,langloishealthy,sebastianik1,saitou,sepangi,ijjas,oikonomouschwarzschild,
kopp,sebastianik2,sebastianik3,chamseddinequanta,rajpoot,odintsovoscillations,
kimura,viablemimeticnoo,mihu,oikonomouaspectsmpla,guendelman5,continuousmedia,focusing,matsumotodescription}. For this reason, we believe it is timely to present a review on the progress achieved thus far in the field of mimetic gravity. This review is not intended to provide a detailed pedagogical introduction to mimetic gravity, rather summarize the main findings and directions in current research on the theory, as well as providing useful directions to the reader, should she/he wish to deepen a given topic in mimetic gravity. For this reason, this review should not be seen as a complete introduction to mimetic gravity, nor should it substitute consultation of the original papers. No prior knowledge on the subject is assumed. \\

This review is structured as follows: in this section we will provide a historical and technical introduction to modified gravity, which shall justify our subsequent endeavour in mimetic gravity. In Section {\bf 2}, we will introduce mimetic gravity. Given that understanding the reason behind the equations of motion of mimetic gravity differing from those of General Relativity is not obvious, a major goal of the Section will be to provide a clear and concise explanation for this fact. Subsequently, in Section {\bf 3} we will explore a few solutions of mimetic gravity, and correspondingly some extensions of the theory, as well as correspondences with related theories of modified gravity. Section {\bf 4} will provide a brief interlude focusing on perturbations in mimetic gravity. In Section {\bf 5} we will present a case study of a mimetic-like model, namely mimetic covariant Ho\v{r}ava-like gravity, with a focus on its solutions and cosmological perturbations, and the need to extend the model beyond its basic formulation. Section {\bf 6} will be devoted to studying spherically symmetric solutions in mimetic gravity. In Section {\bf 7} we will touch upon the issue of rotation curves in mimetic gravity, and how this issue is addressed. Finally, we will conclude in Section {\bf 8}.

\subsection{Why modify General Relativity?}

General Relativity (GR henceforth), first formulated by Albert Einstein in 1915 \cite{einstein1,einstein2,einstein3,einstein4,einstein5} (for a pedagogical review, see e.g. \cite{carroll}), is an extremely successful and predictive theory, and together with Quantum Field Theory forms one of the pillars of modern physics. The traditional picture of GR is a geometrical one, with the theory being one of spacetime and its metric. A more modern view is free of geometrical concepts, and sees GR as the unique theory of massless spin-2 particles. \\

Confirmations of GR abound (see e.g. \cite{will} for a complete review), ranging from gravitational lensing \cite{dyson} to the precession of Mercury's orbit \cite{einstein4}. Shortly after its centennial in 2015 one of the pillars of GR, the existence of gravitational waves, was grandiosely confirmed by the detections of GW150914 \cite{gw150914} and GW151226 \cite{gw151226} by LIGO (see also \cite{corda}). Before we even embark on a review of mimetic gravity, then, it is worth reminding the reader why one should even consider questioning a theory as successful as GR. Aside from the philosophical perspective that questioning theories and exploring other approaches is a sensible route in science, provided of course there is agreement with observations, hints persist in the literature that complicating the gravitational action may indeed have its merits. In fact, the reader should be reminded that as early as 1919 (four years after GR had been formulated), proposals started to be put forward as to how to extend this theory. Notably, in the form of Weyl's scale independent theory \cite{weyl} and Eddington's theory of connections \cite{eddington}. These early attempts to modify GR were driven solely by scientific curiosity with no formal theoretical, or let alone experimental, motivation. \\

Nonetheless, theoretical motivation for modifying the gravitational action came quite soon. The underlying reason being that GR is non-renormalizable and thus not quantizable in the way conventional Quantum Field Theories are quantized. However, it was proven that 1-loop renormalization requires the addition of higher order curvature terms to the Einstein-Hilbert action. In fact, it was later demonstrated that, while actions constructed from invariants quadratic in curvature are renormalizable \cite{stelle}, the addition of higher order time derivatives which follows from the addition of terms higher order in the curvature leads to the appearance of ghost degrees of freedom, which entail a loss of unitarity. More recent results show that when quantum corrections or string theory are taken into account, the effective low energy gravitational action admits higher order curvature invariants (see e.g. \cite{bos} for a general review). However, all these early attempts to modify GR had a common denominator in the fact that terms which modified the gravitational action were only considered to be relevant in proximity of the Planck scale, thus not affecting the late Universe. \\

With the emergence of the ``dark Universe'' picture in recent years, the limits of GR have been fully exposed, and further motivations to modify this theory have emerged. A series of experiments and surveys, including but not limited to CMB experiments, galaxy redshift surveys, cluster surveys, supernovae surveys, lensing experiments, quasar surveys, have depicted a peculiar picture of our Universe \cite{du1,du2,du3,du4,du5,du6,du7,du8,du9,du10,du11,du12,du13,du14,du15,du16,du17,du18,du19,du20,
du21,du22,du23,du24,du25,du26,du27,du28,du29,du30,du31,du32,du33,du34,du35,du36,du37,du38,du39,
du40,du41,du42,du43,du44,du45,du46,du47,du48,du49,du50}. This scenario suggests that our n\"{a}ive picture of the world we live in being described by the Standard Model of Particle Physics (SM) supplemented by General Relativity is, at best, incomplete. The concordance cosmology model suggests a scenario where only $\sim 4\%$ of the energy budget of the Universe consists of baryonic matter, whereas $\sim 24\%$ consists of non-baryonic dark matter (DM), and the remaining $\sim 76\%$ consists of dark energy (DE). Of the last two extra components, dark matter is (presumably) the one with properties most similar to ordinary matter. It shares the clustering properties of ordinary matter, but not its couplings to SM gauge bosons (e.g. electromagnetic ones), and is believed to be responsible for structure formation during the matter-domination era of the cosmological history. As ordinary matter, DM satisfies the strong energy condition. Dark energy, instead, is more peculiar still, given that it does not share the clustering properties of ordinary matter or DM, as it violates the strong energy condition. It it believed to be responsible for the late-time speed-up of the Universe, which has been inferred from a variety of cosmological and astrophysical observations, ranging from type Ia-supernovae to the CMB. Whereas evidence for DE is somewhat indirect and exclusively of cosmological origin, clues as to the existence of DM are instead present on a wide variety of scales, from cosmological to astrophysical (galactic and subgalactic) ones. For technical reviews on DM, see e.g. \cite{dmreview1,dmreview2,dmreview3,dmreview4,dmreview5}, for similar reviews on DE, see for instance \cite{dereview1,dereview2,dereview3,dereview4,dereview5}. \\

The late-time acceleration of our Universe, however, is most likely not the only period of accelerated expansion that our Universe has experienced. A period of accelerated (exponential) expansion during the very early Universe, prior to the conventional radiation and matter domination epochs, is required to solve the horizon, flatness, and monopole problems. This period of accelerated expansion is known as inflation (see e.g. \cite{inf1,inf2,inf3,inf4,inf5,inf6,inf7,inf8,inf9,inf10} for pioneering work), and a vast class of models attempting to reproduce such period exists in the literature (for an incomplete list, see for instance \cite{infmod1,infmod2,infmod3,infmod4,infmod5,infmod6,infmod7,infmod8,infmod9,infmod10} and references therein, see also e.g. \cite{ir1,ir2,ir3,ir4,ir5,ir6,ir7,ir8,ir9,ir10,ir11,ir12,ir13,ir14,ir15,ir16,ir17,ir18,ir19,ir20,ir21,ir22,ir23,ir24,ir25} for more recent inflationary model-building which is extremely closely relevant to mimetic gravity and variations of it). For reviews on inflation, see e.g. \cite{inflationreview1,inflationreview2,inflationreview3,inflationreview4,inflationreview5}. Inflation also purports to be the mechanism generating primordial inhomogeneities which are quantum in origin \cite{perturbations1,perturbations2}, providing the seeds which grow under gravitational instability to form the large-scale structure of the Universe. The fact that the Universe presumably undergoes acceleration at both early and late times or, equivalently, at high and low curvature, is very puzzling, and might be hinting to a more profound structure. \\

It thus appears that concordance cosmology requires at least three extra (possibly dark) cosmological components: one or more dark matter components, some form of dark energy, and one or more inflaton fields. There is no shortage of ideas as to what might be the nature of each of these components. Nonetheless, adding these three or more components opens another set of questions, which include but are not limited to the compatibility with the current SM, and the consistency of formulation. On the other hand, gravity is the least understood of the four fundamental interactions, and the most relevant one on cosmological and astrophysical scales. If so, it could be that our understanding of gravity on these scales is inadequate or incomplete, and modifying our theory of gravitation could indeed be the answer to the dark components of the Universe. One could argue that this solution is indeed more economical and possibly the one to pursue in the spirit of Occam's razor. In other words, modifications to Einstein's theory of General Relativity might provide a consistent description of early and late-time acceleration and of the dark matter which appears to pervade the Universe. Modified theories of gravity not only can provide a solution to the ``dark Universe riddle'', but also possess a number of alluring features such as unification of the various epochs of acceleration and deceleration (matter domination) of the Universe's evolution, transition from non-phantom to phantom phase being transient (and thus without Big Rip), solution to the coincidence problem, and also interesting connections to string theory. \\

Having presented some motivation to modify our theory of gravitation, we now proceed to briefly discuss systematic ways by means of which this purpose can be achieved.

\subsection{How to modify General Relativity?}

Essentially all attempts to modify General Relativity are guided by Lovelock's theorem \cite{lovelock}. Lovelock's theorem states that the only possible second order Euler-Lagrange expression obtainable in 4D space from a scalar Lagrangian density of the form ${\cal L} = {\cal L}(g_{\mu \nu})$, where $g_{\mu \nu}$ is the metric tensor, is given by the following:
\begin{eqnarray}
E ^{\mu \nu} = \beta \sqrt{-g} \left ( R ^{\mu \nu} - \frac{1}{2}g ^{\mu \nu}R \right ) + \kappa \sqrt{-g}g ^{\mu \nu} \, ,
\end{eqnarray}
$\beta$ and $\kappa$ being constants, and $R ^{\mu \nu}$ and $R$ being the Ricci tensor and scalar respectively. It follows that constructing metric theories of gravity whose equations differ from those of GR requires at least one of the following to be satisfied:
\begin{itemize}
\item Presence of other fields apart from or \textit{in lieu} of the metric tensor
\item Work in a number of dimensions different from 4
\item Accept metric derivatives of degree higher than 2 in the field equations
\item Give up locality or Lorentz invariance
\end{itemize}
Therefore, we can imagine broadly classifying the plethora of modified gravity theories according to which of the above assumptions is broken.

\subsubsection*{New degrees of freedom}

Relaxing the first assumption leads to what is probably the largest class of modified gravity theories. Theories corresponding to the addition of \textit{scalar} degrees of freedom include quintessence (e.g. \cite{quintessence1,quintessence2,quintessence3}) and coupled quintessence (e.g. \cite{coupledquintessence}) theories, the Chern-Simons theory (e.g. \cite{chernsimons}), Cuscuton cosmology (e.g. \cite{cuscuton1,cuscuton2,cuscuton3}), Chaplygin gases (e.g. \cite{chaplygin1,chaplygin2}), torsion theories such as f(T) theories (e.g. \cite{ft1,ft2,ft3,ft4}, see also e.g. \cite{fto1,fto2,fto3,fto4,fto5,fto6,fto7,fto8,fto9,fto10,fto11} for recent work) or the Einstein-Cartan-Sciama-Kibble theory (e.g. \cite{ecsk1,ecsk2,ecsk3,ecsk4}), scalar-tensor theories (e.g. \cite{st}) such as the Brans-Dicke theory \cite{bransdicke}, ghost condensates (e.g. \cite{ghostcondensates}), galileons (e.g. \cite{galileon1,galileon2}), KGB \cite{kgb}, Horndeski's theory \cite{horndeski}, and many others. \\

One can instead choose to add \textit{vector} degrees of freedom, as in the case of the Einstein-Aether theory (e.g. \cite{ea1,ea2,ea3,ea4,ea5}). The addition of a vector field leads to the introduction of a preferred direction in spacetime, which entails breaking Lorentz invariance. \\

Theories where \textit{tensor} degrees of freedom are added include Eddington-Born-Infeld gravity (e.g. \cite{ebi1,ebi2,ebi3,ebi4}), and bimetric MOND (e.g. \cite{bimond1,bimond2}) among others. TeVeS (Tensor Vector Scalar gravity, \cite{tvs}) is instead a theory which features the addition of all three types of degrees of freedom together. \\

Broadly speaking, mimetic gravity belongs to the class of theories of modified gravity where an additional scalar degree of freedom is added. Caution is needed with this identification though because, as we shall see later, mimetic gravity does not possess a proper scalar degree of freedom, but rather a constrained one. \\

\subsubsection*{Extra dimensions}

Relaxing the second assumption instead brings us to consider models with extra dimensions, the prototype of which is constituted by Kaluza-Klein models (e.g. \cite{kaluzaklein1,kaluzaklein2,kaluzaklein3}). Models of modified gravity with extra dimensions abound when considering string theory, including Randall-Sundrum I \cite{randallsundrum1} and II \cite{randallsundrum2} models, Einstein-Dilaton-Gauss-Bonnett gravity (e.g. \cite{edgb1,edgb2}), cascading gravity (e.g. \cite{cascading1,cascading2,cascading3,cascading4}), Dvali-Gabadadze-Porrati gravity (e.g. \cite{dgp1,dgp2}). Another interesting theory that falls within the extra dimension category is represented by 2T gravity \cite{2t}.

\subsubsection*{Higher order}

The most famous and studied example of a theory falling within this category is undoubtedly represented by $f(R)$ gravity (\cite{fr1}, see also e.g. \cite{nojiriodintsov1,nojiriodintsov3,clifton,delacapozziello,fr2,fr3,fr4,fr5,fr6,fr7,fr8} or \cite{davoodfR} for black holes phenomenology). In fact, unification of inflation and late-time acceleration was proposed in the context of $f(R)$ gravity in \cite{fr2,fr4,odintsovadd1,odintsovadd2,odintsovadd3}. Belonging to the same family are also variations of the former such as $f(R_{\mu \nu}R^{\mu \nu})$, $f(\Box R)$, $f(R,T)$ $f(R, T, R^{\mu \nu}T_{\mu \nu})$ gravity (see e.g.\cite{higherorder1,higherorder2,higherorder3,higherorder4,higherorder5,higherorder6,
higherorder7,higherorder8,higherorder9,higherorder10,higherorder11}), but also Gauss-Bonnet (see e.g. \cite{gb1,gb2,gb3,gb4,gb5,gb6,gb7,gb8,gb9,gb10,gb11}) and conformal gravity (see e.g. \cite{conformal1,conformal2,conformal3}). Another well-known theory which lies within this family is represented by Ho\v{r}ava-Lifshitz gravity (\cite{horava}, see also \cite{horava1,horava2,horava3,horava4,horava5,horava6,horava7,horava8,horava9,horava10,horava11,horava12,
horava13,horava14,horava15,horava16,horava17,horava18,horava19,horava20,horava21,horava22,horava23,horava24,horava25} , which also violates Lorentz invariance explicitly), and correspondingly the vast category of Ho\u{r}ava-like theories, including those which break Lorentz invariance dynamically, e.g. \cite{horava18,horava23,horava24,crg1,crg2,crg3,crg4,crg5,crg6,crg7,crg8,crg9}.

\subsubsection*{Non-local}

If we choose to relax the assumption of locality (we have already seen cases where the assumption of Lorentz invariance is relaxed above), we can consider non-local gravity models whose action contains the inverse of differential operators of curvature invariants, such as $f(R/\Box)$ and $f(R_{\mu \nu}\Box ^{-1}R ^{\mu \nu})$ gravity (e.g. \cite{deserwoodard,nojirinonlocal,thongkool,nojiriodintsovsasakizhang,vernov}). Some degravitation scenarios belong to this family as well, e.g. \cite{modesto,jaccardmaggioremitsou}. \\

Broadly speaking, mimetic gravity belongs to the class of theories of modified gravity where an additional scalar degree of freedom is added. Caution is needed with this identification though because mimetic gravity does not possess a proper scalar degree of freedom. Instead, the would-be scalar degree of freedom is constrained by a Lagrange multiplier, which kills all higher derivatives. As such, the mimetic field cannot have oscillating solutions and the sound speed satisfies $c_s = 0$, confirming that there is no propagation of scalar degrees of freedom (however, this is true only in the original mimetic model, but does not necessarily hold in extensions thereof). Furthermore, the same Lagrange multiplier term introduced a preferred foliation of spacetime, which breaks Lorentz invariance (although this is preserved at the level of the action). These aspects of mimetic gravity will be discussed in more detail in the following Section. \\

\section{Mimetic gravity\label{MG}}

The expression ``mimetic dark matter'' was first coined in a 2013 paper by Mukhanov and Chamseddine \cite{m1} although, as we shall see later, the foundation for mimetic theories had actually been developed a few years earlier in three independent papers \cite{lim,gao,capozziello}. In \cite{m1}, the proposed idea is to isolate the conformal degree of freedom of gravity by introducing a parametrization of the physical metric $g_{\mu \nu}$ in terms of an auxiliary metric $\tilde{g}_{\mu \nu}$ and a scalar field $\phi$, dubbed mimetic field, as follows:
\begin{eqnarray}
g _{\mu \nu} = -\tilde{g} _{\mu \nu}\tilde{g} ^{\alpha \beta}\partial _{\alpha}\phi\partial _{\beta}\phi \,.
\label{mimeticmetric}
\end{eqnarray}
From Eq.(\ref{mimeticmetric}) it is clear that, in such a way, the physical metric is invariant under conformal transformations of the auxiliary metric of the type $\tilde{g} _{\mu \nu} \rightarrow \Omega(t, {\bf x}) ^2\tilde{g} _{\mu \nu}$, $\Omega(t, {\bf x})$ being a function of the space-time coordinates. It is also clear that, as a consistency condition, the mimetic field must satisfy the following constraint:
\begin{eqnarray}
g ^{\mu \nu}\partial _{\mu}\phi\partial _{\nu}\phi = -1\,.
\label{norm}
\end{eqnarray}
Thus, the gravitational action, taking into account the reparametrization given by Eq.(\ref{mimeticmetric}) now takes the form:
\begin{eqnarray}
I=\frac{1}{2}\int_\mathcal{M}d^4 x\sqrt{-g\left(\tilde g_{\mu\nu},\phi\right)}\left[R(\tilde g_{\mu\nu}, \phi)+2\mathcal L_m\right]\,,
\label{mimeticaction}
\end{eqnarray}
where $\mathcal M$ is the space-time manifold, $R\equiv R(g_{\mu\nu},\phi)$ is the Ricci scalar, $\mathcal L_m$ is the matter Lagrangian and $g\equiv g(\tilde g_{\mu\nu},\phi)$ is the determinant of the physical metric. \\

By varying the action with respect to the physical metric one obtains the equations for the gravitational field. However, this process must be done with care, for the (variation of the) physical metric can be written in terms of the (variation of the) auxiliary metric and the (variation of the) mimetic field. Taking this dependency into account, variation of the action with respect to the physical metric yields \cite{m1}:
\begin{eqnarray}
G _{\mu \nu} - T _{\mu \nu} + (G-T)\partial _{\mu}\phi\partial _{\nu}\phi = 0 \,,
\label{basic}
\end{eqnarray}
where $G_{\mu\nu}=R_{\mu\nu}-g_{\mu\nu}R/2$ is the Einstein tensor, $R_{\mu\nu}$ being the Ricci tensor, while $G(=-R)$ and $T$ are the trace of the Einstein's tensor and the stress energy tensor of matter, respectively. Notice that the auxiliary metric does not enter the gravitational field equation explicitly, but only through the physical metric, whereas the mimetic field enters explicitly. In fact, the mimetic field contributes to the right hand side of Einstein's equation through an additional stress-energy tensor component:
\begin{eqnarray}
\tilde T_{\mu\nu}=-(G-T)\partial _{\mu}\phi\partial _{\nu}\phi\,.
\label{tensorphi}
\end{eqnarray}
We note that both energy-momentum tensors, $T_{\mu\nu}$ and $\tilde T_{\mu\nu}$, are covariantly conserved, i.e. $\nabla^\mu T_{\mu\nu}=\nabla^\mu \tilde T_{\mu\nu}=0$ (with $\nabla^\mu$ the covariant derivative), whereas the continuity equation for $\tilde T_{\mu\nu}$ with the mimetic constraint Eq.(\ref{norm}) leads to:
\begin{eqnarray}
\nabla^\kappa\left((G-T)\partial_\kappa\phi\right)\equiv\frac{1}{\sqrt{-g}}\partial_\kappa\left(\sqrt{-g}(G-T)g^{\kappa\sigma}\partial_\sigma\phi\right)=0\,.
\label{phieq}
\end{eqnarray}
Finally, the trace of Eq.(\ref{basic}) is found to be:
\begin{eqnarray}
(G-T)(1+g^{\mu\nu}\partial_\mu\phi\partial_\nu\phi)=0\,.
\end{eqnarray}
It is clear that the above is automatically satisfied if one takes into account Eq.(\ref{norm}) even if $G\neq T$. Thus the theory admits non-trivial solutions and the conformal degree of freedom becomes dynamical even in the absence of matter ($T=0$ but $G\neq 0$) \cite{m1}. \\

Let us examine the structure of the mimetic stress-energy tensor. Recall that the stress-energy tensor of a perfect fluid whose energy density is $\rho$ and pressure $p$ is given by:
\begin{eqnarray}
T_{\mu\nu}=(\rho+p)u_\mu u_\nu+p g_{\mu\nu}\,,\quad u_\mu u^\mu=-1\,,
\label{stresspf}
\end{eqnarray}
Notice that the mimetic stress-energy tensor in  Eq.(\ref{tensorphi}) assumes the same form of the one of a perfect fluid with pressure $p=0$ and  energy density $\rho=-(G-T)$, while the gradient of the mimetic field, $\partial _{\mu}\phi$, plays the role of 4-velocity. Thus the mimetic fluid is pressureless, suggesting it could play the role of dust in a cosmological setting. To confirm whether the mimetic fluid can indeed play the role of dust, it is necessary to investigate cosmological solutions. In fact, this is easy to do on a Friedmann-Lema\^{i}tre-Robertson-Walker (FLRW) setting, with a metric of the form:
\begin{eqnarray}
ds^2 = -d t^2 +a(t)^2 d{\bf x}^2 \, ,\label{FRWmetric}
\end{eqnarray}
where $a\equiv a(t)$ is the scale factor. If we take the hypersurfaces of constant time to be equal to those of constant $\phi$, we immediately see that the constraint equation, Eq.(\ref{norm}), is automatically satisfied if the mimetic field is identified with time up to an integration constant (which we arbitrarily set to 0). Thus, the mimetic field plays the role of ``clock'' on an FLRW background. It is then easy to show that Eq.(\ref{phieq}) implies that $(G - T)$, which corresponds to the energy density of the mimetic stress-energy tensor, decays with the scale factor of the FLRW Universe as $(G - T) \propto 1/a^3$. Recall that the energy density of a component with equation of constant state parameter $w$ evolves as $a^{-3(w+1)}$ in an FLRW Universe, and hence the evolution in the energy density of the mimetic field corresponds to $w = 0$, namely the equation of state for dust. In other words, the conformal degree of freedom of gravity can mimic the behaviour of dark matter at a cosmological level, hence the name ``mimetic dark matter'' \cite{m1}. \\

\subsubsection{Lagrange multiplier formulation}

Before further discussing some fundamental aspects of mimetic dark matter (or mimetic gravity henceforth), such as the reason behind the different solutions from GR despite the seemingly innocuous parametrization given by Eq.(\ref{mimeticmetric}), let us discuss an alternative but equivalent formulation of mimetic gravity. The equations of motion obtained from the action written in terms of the auxiliary metric $\tilde{g}$ are equivalent to those one would conventionally obtain from the action expressed in terms of the physical metric with the imposition of an additional constraint on the mimetic field. This suggests that the mimetic constraint given by Eq.(\ref{norm}) can actually be implemented at the level of the action by using a Lagrange multiplier. That is, the action for mimetic gravity, Eq.(\ref{mimeticaction}), can be written as:
\begin{eqnarray}
I=\frac{1}{2}\int_\mathcal M d^4 x\sqrt{-g}\left[R+\lambda(g^{\mu\nu}\partial_\mu\phi\partial_\nu\phi+1)+2\mathcal L_m\right]\,.
\label{actionlm}
\end{eqnarray}
Variation of the Lagrangian respect to the Lagrange multiplier field $\lambda$ leads to Eq.(\ref{norm}), while variation respect to the physical metric $g_{\mu\nu}$ yields,
\begin{eqnarray}
G_{\mu\nu}-T_{\mu\nu}+\lambda\partial_\mu\phi\partial_\nu\phi=0\,,
\end{eqnarray}
whose trace, when one takes into account Eq.(\ref{norm}), is given by,
\begin{eqnarray}
\lambda= (G-T)\,.
\end{eqnarray}
Thus, one recovers Eq.(\ref{basic}) again. In this review, we will always make use of the action given by Eq.(\ref{actionlm}) to introduce the mimetic field. \\

A remark is in order here. Actions such as Eq.(\ref{actionlm}) had actually been introduced three years before the term ``mimetic dark matter'' was first coined. Three independent papers in 2010, by Lim, Sawicki, \& Vikman \cite{lim}, Gao, Gong, Wang, \& Chen \cite{gao}, and Capozziello, Matsumoto, Nojiri, \& Odintsov \cite{capozziello} respectively, presented models with two additional scalar fields, one of which playing the role of Lagrange multiplier enforcing a constraint on the derivative of the other.\footnote{See also \cite{paliathanasis} for recent work on the role of Lagrange multiplier constrained terms in cosmology.} In fact, it was shown that these types of models can produce an unified theory describing dark energy and dark matter, because the term inside the Lagrange multiplier can always be arranged in such a way to reproduce the conventional expansion history of $\Lambda$CDM. Thus it is fair to state that mimetic theories had actually seen birth prior to the 2013 paper by Mukhanov and Chamseddine.

\subsection{Understanding mimetic gravity}

Before we can make further progress in exploring solutions in mimetic gravity, generalizing the theory, or studying connections to other theories, we need to touch on two very important points: first, why the seemingly innocuous parametrization given by Eq.(\ref{basic}) has led to a completely new class of solutions not contemplated by GR. And second, whether the theory is stable or not. As we shall see, the first point can ultimately be explained in terms of singular disformal transformations. \\

It might appear puzzling at first that, only by rearranging parts of the metric, one is faced with a different model altogether. A first explanation appeared in \cite{golovnev}, which explained this property in terms of variation of the action taking place over a restricted class of functions. This is the case in mimetic gravity, precisely because the consistency equation Eq.(\ref{norm}) enforces an additional condition for any admissible variation of the action, in particular demanding that:
\begin{eqnarray}
\int _{t_{\text{in}}} ^{t_{\text{fin}}} dt \sqrt{\Omega(x)} = (t_{\text{fin}} - t_{\text{in}} )\quad , \quad \Omega (x) \equiv \tilde{g}^{\alpha \beta}\partial _{\alpha}\phi\partial _{\beta}\phi \, .
\end{eqnarray}
Thus for a spatially homogeneous mimetic field $\phi = t$, $\dot{\phi} = \sqrt{\Omega}$. Varying over a restricted class of functions now provides less conditions for the stationarity of the action, and hence more freedom in the dynamics. This is a well known property relevant when one makes derivative substitutions $x \equiv f(\dot{y})$ into an action $I(x,t)$: the class of functions over which the variation is admissible does not only comprise those for which the variation of $x$ is vanishing at the boundary, but also those such that the integral of the variation of $x$ is zero. This extra restriction leads to less conditions for stationarity of the action and the appearance of additional dynamics with respect to the original case. \\

Another explanation was presented in \cite{barvinsky}, which identified mimetic gravity as a conformal (Weyl-symmetric) extension of GR. The first important point to notice is that the parametrization $g ^{\mu \nu} = g^{\mu \nu}(\tilde{g} ^{\mu \nu}, \phi)$ is non-invertible even for fixed $\phi$, owing to the fact that the map $g \rightarrow \tilde{g}, \phi$ is a map from ten variables to eleven. With this parametrization, the theory is manifestly conformally invariant, i.e. invariant with respect to transformations of the auxiliary metric (and correspondingly the action $I$) of the form:
\begin{eqnarray}
\partial_{\alpha}\tilde{g}_{\mu \nu}(x) = \alpha(x)g_{\mu \nu}(x) \quad , \quad \partial_{\alpha}I\left [ g_{\mu \nu}(\tilde{g}_{\mu \nu},\phi) \right ] = 0 \,,
\end{eqnarray}
where $\alpha(x)$ is a function of the space-time coordinates.
Two immediate corollaries of the theory's conformal invariance are its yielding identically traceless equations of motion for the gravitational field, and its requiring conformal gauge fixing. In fact, recall that the equation of motion for the gravitational field, Eq.(\ref{basic}), is automatically traceless if the consistency condition given by Eq.(\ref{norm}) is satisfied. Therefore, we can identify Eq.(\ref{norm}) with the conformal gauge condition in the locally gauge-invariant theory with action $I \left [ g_{\mu \nu} ( \tilde{g}_{\mu \nu},\phi) \right]$. Mimetic gravity can thus be seen as a conformal extension of GR which is Weyl-invariant in terms of the auxiliary metric $\tilde{g}_{\mu \nu}$. However, this theory is quite different from the off-shell conformal extensions of GR proposed in \cite{fradkin} (which preserves GR on-shell but modifies its effective action off-shell), here the gravitational action is already modified at a classical level, by adding an extra degree of freedom provided by a collisionless perfect fluid. This additional degree of freedom, which can mimic collisionless cold dark matter for cosmological purposes, arises from gauging out local Weyl invariance. \\

\subsubsection{Singular disformal transformations}

As we anticipated above, mimetic gravity and the appearance of the extra degree of freedom which can mimic cosmological dark matter, are rooted into the role played by singular disformal transformations. As was shown by Bekenstein \cite{bekenstein}, because GR enjoys invariance under diffeomorphisms, one is free to parametrize a metric $g_{\mu \nu}$ in terms of a fiducial metric $\tilde g_{\mu \nu}$ and a scalar field $\phi$. The map between the two is defined as a ``disformal transformation'', or a ``disformation'', and takes the following form:
\begin{eqnarray}
g_{\mu \nu} = {\cal A}(\phi, X)\tilde{g}_{\mu \nu} + {\cal B}(\phi , X)\partial_{\mu}\phi\partial_{\nu}\phi \, ,
\label{dis}
\end{eqnarray}
where $X \equiv \tilde{g}^{\mu \nu}\partial_{\mu}\phi\partial_{\nu}\phi$. The functions ${\cal A}$ and ${\cal B}$ are referred to as conformal factor and disformal factor, respectively. In general the functions ${\cal A}(\phi , X)$, ${\cal B}(\phi , X)$ are arbitrary, with ${\cal A} \neq 0$. It is easy to show that, provided the transformation is invertible, the equations of motion for the theory (obtained by variation of the action with respect to $\tilde{g}_{\mu \nu}$ and $\phi$) reduce to those of obtained by varying with respect to the metric $g_{\mu \nu}$ \cite{deruelle}. \\

To make progress, it is useful to contract the two equations of motion with $\tilde{g}_{\mu \nu}$ and $\partial_{\mu}\phi\partial_{\nu}\phi$. Although we will not perform the steps explicitly, it is easy to show that this leads to the following two equations of motion:
\begin{eqnarray}
\Omega \left ( {\cal A} - X\frac{\partial {\cal A}}{\partial X} \right ) - \omega X\frac{\partial {\cal B}}{\partial X} = 0 \quad , \quad \Omega X^2\frac{\partial {\cal A}}{\partial X} - \omega \left ( {\cal A} - X^2\frac{\partial {\cal B}}{\partial X} \right ) = 0 \, ,
\label{sys}
\end{eqnarray}
where the two quantities $\Omega$ and $\omega$ are defined as:
\begin{eqnarray}
\Omega \equiv (G^{\mu \nu} - T^{\mu \nu})\tilde{g}_{\mu \nu} \quad , \quad \omega \equiv (G^{\mu \nu} - T^{\mu \nu})\partial_{\mu}\phi\partial_{\nu}\phi \, .
\end{eqnarray}
The determinant of the system Eq.(\ref{sys}) is given by:
\begin{eqnarray}
det = X^2{\cal A}\frac{\partial}{\partial X} \left ( {\cal B} + \frac{{\cal A}}{X} \right ) \, .
\label{determinant}
\end{eqnarray}
If the above is nonzero, it is trivial to obtain that the resulting set of equations consists of Einstein's equation ($G_{\mu \nu} = T_{\mu \nu}$) and a second empty equation. Therefore, the theory does not feature new solutions with respect to GR \cite{deruelle}. \\

The situation is quite different if the determinant in Eq.(\ref{determinant}) is zero, which corresponds to the physical case when the disformal transformation given by Eq.(\ref{dis}) is non-invertible, or singular. In this case, being the function ${\cal A} \neq 0$, this immediately determines ${\cal B}$, which is of the form:
\begin{eqnarray}
{\cal B}(X , \phi) = -\frac{{\cal A}(X , \phi)}{X} + {\cal E}(\phi) \, ,
\label{dt}
\end{eqnarray}
where ${\cal E}(\phi) \neq 0$ is an arbitrary function. We will not show the steps explicitly, which are instead discussed in detail in Section IV. of \cite{deruelle}, but it is not hard to obtain the equations of motion and notice that they differ from those of GR, due to the presence of an extra term on the right-hand side of Einstein's equation (i.e. an additional contribution to the stress-energy tensor). Therefore, when the disformal transformation is singular, one is faced with the appearance of extra degrees of freedom which result in equations of motion differing from those of GR \cite{deruelle}. \\

The parametrization Eq.(\ref{basic}) defining mimetic gravity, can be identified with a singular disformal transformation, with ${\cal A} = X$ and ${\cal B} = 0$ in Eq.(\ref{dis}), and correspondingly ${\cal E} = 1$ in Eq.(\ref{dt}). In general, when the relation defined by Eq.(\ref{dt}) exists between the conformal factor ${\cal A}$ and the disformal factor ${\cal B}$, the resulting disformal transformation is singular and, as a result, the system possesses additional degrees of freedom, explaining the origin of the extra degree of freedom in mimetic gravity which mimics a dust component. This aspect has been at the center of a number of studies recently, see e.g. \cite{deruelle,yuan,domenech,deffayet,arroja,mukohyama,bartolo,langlois,langloishealthy,carvalho,ghalee1,ghalee2,ghalee3}. Moreover, \cite{arroja} has shown that the two approaches towards mimetic gravity (and further extensions to be discussed later, such as mimetic Horndeski theories), namely Lagrange multiplier [Eq.(\ref{actionlm})] and singular disformal transformation [Eq.(\ref{dis})] are in fact equivalent. \\

\subsubsection{Mimetic gravity from the Brans-Dicke theory}

There actually exists a third route to mimetic gravity, apart from disformal transformations and Lagrange multiplier, whose starting point is the singular Brans-Dicke theory. Namely, by starting from the action Eq.(\ref{mimeticaction}) (neglecting matter terms), and by performing the conformal transformation given by Eq.(\ref{basic}), we end up with the action \cite{hammer}:
\begin{eqnarray}
I(\tilde{g},\phi) = \int d^4x \sqrt{-\tilde{g}} \left ( XR(\tilde{g}) + \frac{3}{2}\frac{\tilde{g}^{\mu\nu}\partial_{\mu}X\partial_{\nu}X}{X} \right ) \, ,
\label{bda}
\end{eqnarray}
where we have defined $X \equiv \tilde{g}^{\mu\nu}\partial_{\mu}\phi\partial_{\nu}\phi/2$. One immediately sees that Eq.(\ref{bda}) corresponds to the singular/conformal Brans-Dicke action \cite{bransdicke} with density parameter $\omega = -3/2$. Thus, we conclude that a third way of obtaining mimetic gravity is by substituting the kinetic term \textit{in lieu} of the scalar field in the singular Brans-Dicke action \cite{hammer}. In case matter fields are included, the substitution has to be performed on the matter part of the Lagrangian as well, which means that matter will not be coupled to $\tilde{g}_{\mu\nu}$ but to $2X\tilde{g}_{\mu\nu}$ \cite{hammer}.

\subsection{Stability}

Is mimetic gravity stable? In other words, does its spectrum contemplate the presence of states with negative norm, or fields whose kinetic term has the wrong sign (corresponding to negative energy states), which could possibly destabilize the theory? This is an important question which has yet to find a definitive answer. Recall that the original mimetic theories formulated in 2010 were found to suffer from a tachyonic instability \cite{kluson}, therefore the question of whether mimetic gravity is stable is totally pertinent. \\

If we formulate the theory of mimetic gravity using the physical metric $g_{\mu \nu}$, we inevitably incur into the risk of the appearance of higher derivatives of the mimetic field, which could entail the emergence of ghosts. Addressing this question requires performing a Hamiltonian analysis of the theory, identifying all constraints and counting the local degrees of freedom. A preliminary analysis of this problem was conducted in \cite{barvinsky}, which concluded that the theory is stable if the energy density of the mimetic field $\epsilon = T - G = T + R$ is positive. This condition is, of course, easy to understand physically. Moreover, it indicates a preference for de Sitter-type backgrounds with a positive cosmological constant, since in that case both contributions to the energy density, given by curvature and trace of the matter stress-energy tensor, are positive. Therefore, it is presumed that mimetic is gravity is stable provided the time evolution of the system preserves the positivity of the energy density stored in the mimetic degree of freedom. The work of \cite{barvinsky} also identified another possible instability issue, namely caustic singularities (which are not dangerous at the quantum level, unlike ghost instabilities). These are presumably due to the pressureless nature of the mimetic field and can possibly be circumvented if one modifies the theory with higher derivative terms, as we will discuss later. \\

The analysis of \cite{barvinsky} imposed the conformal gauge condition Eq.(\ref{norm}) prior to proceeding to the canonical formulation. A proper analysis should instead take place in full generality, and has been conducted in \cite{chaichian}. This analysis finds that the Hamiltonian constraint depends linearly on the momentum, which in most cases signals that the Hamiltonian density of the theory is unbounded from below, a classical sign of instability. As anticipated, this occurs frequently for higher derivative theories, which are prone to the Ostrogradski instability. The work in \cite{chaichian} concluded, as \cite{barvinsky}, that mimetic gravity is stable as long as the energy density in the dust degree of freedom in the theory remains positive. However, this is not always consistent with the dynamics of the theory, given that for some initial configurations, the energy density could start its evolution with a positive value but then end up with a negative value, which would cause instability. In fact, \cite{chaichian} finds that the requirement that the theory be stable correspond to the requirement that initial configurations do not cross the surface for which the momentum conjugate to the mimetic field, $p_{\phi}$, satisfies $p_{\phi} = 0$. \\

A possible solution to these instability issues was presented in \cite{barvinsky} and studied in \cite{chaichian}. The idea is to modify the parametrization Eq.(\ref{basic}) by making use, instead of the gradient of a scalar field, of a dynamical vector (Proca) field:
\begin{eqnarray}
g _{\mu \nu} = -\tilde{g} _{\mu \nu}\tilde{g} ^{\alpha \beta}u_{\alpha}u_{\beta} \,.
\label{proca}
\end{eqnarray}
The Proca field is made dynamical by adding a Maxwell kinetic term $F^2$ to the action, where $F$ is the field-strength tensor of the vector field. It is beyond the scope of our review to provide details of the analysis, conducted in \cite{chaichian} which finds that the Hamiltonian in the Proca mimetic model shows no sign of instability. Furthermore, \cite{chaichian} proposes an interesting extension of mimetic gravity to a mimetic tensor-vector-scalar model, by generalizing Eq.(\ref{mimeticmetric}) to:
\begin{eqnarray}
g _{\mu \nu} = -f(\phi)\tilde{g} _{\mu \nu}\tilde{g} ^{\alpha \beta}u_{\alpha}u_{\beta} \,,
\end{eqnarray}
where now both the scalar and vector degrees of freedom contribute to mimetic matter. It is furthermore demonstrated that the theory is free of ghosts \cite{chaichian}. \\

Another recent work confirmed in all generality that the original mimetic gravity theory suffers from ghost instability \cite{langlois}, in the following way. It is immediate to show that mimetic gravity is invariant under the local symmetry defined by:
\begin{eqnarray}
\delta \phi = 0 \quad , \quad \delta g_{\mu \nu} = \epsilon \left ( \frac{\partial {\cal A}}{\partial X} g_{\mu \nu} + \frac{\partial {\cal B}}{\partial X}\partial_{\mu}\phi\partial_{\nu}\phi \right ) \, ,
\label{conformal}
\end{eqnarray}
where as usual ${\cal A}$ and ${\cal B}$ correspond to the conformal and disformal factors. Being ${\cal A} = X$ and ${\cal B} = 0$ for mimetic gravity, Eq.(\ref{conformal}) corresponds as expected to invariance of the physical metric under conformal transformations of the auxiliary metric. In the Hamiltonian description, this symmetry is associated to a first class constraint. In fact, one can show that the primary constraint, which corresponds to the generator of infinitesimal conformal transformations, is first class, with its Poisson commuting with the Hamiltonian and momentum constraints. This leaves no place for a secondary constraint which could eliminate the Ostrogradski ghost. Thus, this result confirms indeed that the original mimetic gravity proposal suffers from a ghost instability.

\section{Solutions and extensions of mimetic gravity}

Having discussed the underlying physical foundation of mimetic gravity, and its stability, we can now proceed to study solutions and extensions of this theory. \\

\subsection{Potential for mimetic gravity}

Recall that, in a cosmological setting, the mimetic field plays the role of ``clock''. Therefore, one can imagine making the mimetic field dynamical by adding a potential for such field to the action. A field-dependent potential corresponds to a time-dependent potential which, by virtue of the Friedmann equation, corresponds to a time-varying Hubble parameter (and correspondingly scale factor). Therefore, by adding an appropriate potential for the mimetic field, one can in principle reconstruct any desired expansion history of the Universe. This is the idea behind the minimal extension of mimetic gravity first proposed in \cite{m2}. The action of mimetic gravity (in the Lagrange multiplier formulation) is thus extended to include a potential for the mimetic field:
\begin{eqnarray}
I = \frac{1}{2}\int_\mathcal M d^4 x\sqrt{-g} \left [ R+\lambda(g^{\mu\nu}\partial_\mu\phi\partial_\nu\phi+1) - V(\phi) + 2\mathcal L_m \right ] \, .
\end{eqnarray}
The equations of motion of the theory are then given by:
\begin{eqnarray}
G_{\mu \nu} - 2\lambda \partial_{\mu}\phi\partial_{\nu}\phi -g_{\mu \nu}V(\phi) = T_{\mu \nu} \, ,
\label{potential}
\end{eqnarray}
which, by taking the trace, can be used to determine the Lagrange multiplier:
\begin{eqnarray}
\lambda = \frac{1}{2} \left ( G - T - 4V \right ) \, .
\label{mul}
\end{eqnarray}
When plugged into Eq.(\ref{potential}), Eq.(\ref{mul}) yields:
\begin{eqnarray}
G_{\mu \nu} = (G - T - 4V)\partial_{\mu}\phi\partial_{\nu}\phi + g_{\mu \nu}V(\phi) + T_{\mu \nu} \, .
\end{eqnarray}
Of course, variation with respect to the Lagrange multiplier as usual yields the constraint Eq.(\ref{norm}). Thus, when a potential is added to the action, the mimetic field contributes a pressure and energy density of $p = -V$ and $\rho = G - T - 3V$ respectively \cite{m2}. One further equation of motion can be derived by varying the action with respect to the mimetic field, which gives:
\begin{eqnarray}
\nabla ^{\nu} \left [ \left ( G - T - 4V \right ) \partial_{\nu}\phi \right ] = - \frac{\partial V}{\partial \phi} \, .
\label{equationmimeticfield}
\end{eqnarray}
when taking into account the expression for the Lagrange multiplier Eq.(\ref{mul}). \\

To study cosmological solutions, it is useful to consider a flat FLRW background [Eq.(\ref{FRWmetric})], since therein the mimetic field can be identified with time. In this case, it is not hard to show that Eq.(\ref{equationmimeticfield}) reduces to \cite{m2}:
\begin{eqnarray}
\frac{1}{a^3}\frac{d}{dt} \left [ a^3 \left ( \rho - V \right ) \right ] = - \frac{dV}{dt} \, ,
\end{eqnarray}
which can be integrated to give:
\begin{eqnarray}
\rho = \frac{3}{a^3}\int da \ a^2V \, ,
\label{energy}
\end{eqnarray}
whereas the pressure remains $p = -V$. The Friedmann equation can instead be manipulated to the form:
\begin{eqnarray}
2\dot{H} + 3H^2 = V(t) \, ,
\label{friedmann}
\end{eqnarray}
where as usual the Hubble parameter is defined as $H \equiv \dot{a}/a$. Further progress can be made by performing the substitution $y \equiv a^{3/2}$, which yields the following equation \cite{m2}:
\begin{eqnarray}
\ddot{y} - \frac{3}{4}V(t)y = 0 \, .
\label{yeq}
\end{eqnarray}
It should be noticed that the equations of motion simplify greatly because of the identification of the mimetic field with time on an FLRW background. In this way, the pressure becomes a known function of time and $y$ satisfies a linear  differential equation. We now proceed to study a few interesting potentials and the corresponding solutions. \\

\subsubsection{$\mathbf{V \propto 1/\phi^2}$}

Let us consider the following potential \cite{m2}
\begin{eqnarray}
V (\phi) = \frac{\alpha}{\phi^2} = \frac{\alpha}{t^2} \, .
\end{eqnarray}
Solving the corresponding Eq.(\ref{yeq}) and substituting for the scale factor, $a \equiv y^{2/3}$, yields the following solution for $\alpha \geq -1/3$:
\begin{eqnarray}
a(t) = t^{\frac{1}{3}(1 + \sqrt{1 + 3\alpha})}\left ( 1 + \beta t^{-\sqrt{1+3\alpha}} \right ) ^{\frac{2}{3}} \, ,
\end{eqnarray}
where $\beta$ is an integration constant. For $\alpha < -1/3$ the solution describes an oscillating flat Universe with amplitude of oscillations which grows with time, however the solution presents singularities and for this reason we will not write it down explicitly \cite{m2}. We can furthermore determine the equation of state parameter (EoS) $w \equiv p/\rho$ if we recall that the energy density is given by Eq.(\ref{energy}) whereas the pressure reads $p = -V$. Explicit calculation gives the following:
\begin{eqnarray}
w = -3\alpha \left ( 1 + \sqrt{1 + 3\alpha}\frac{1 - \beta t^{-\sqrt{1 + 3\alpha}}}{1 + \beta t^{-\sqrt{1 + 3\alpha}}} \right ) ^{-2} \, .
\end{eqnarray}
It is interesting to note that, for $\alpha \gg 1$, the EoS approaches $w = -1$, i.e. a cosmological constant, at late times. \\

We can consider the case where mimetic matter is a subdominant energy component in the Universe, which is instead dominated by another form of matter with EoS $\tilde{w}$. The scale factor in this situation evolves as:
\begin{eqnarray}
a \propto t^{\frac{2}{3(1+\tilde{w})}} \, ,
\end{eqnarray}
and hence Eq.(\ref{energy}) can be used to deduce the evolution of the energy density of mimetic matter, which decays as:
\begin{eqnarray}
\rho = -\frac{\alpha}{\tilde{w}t^2} \, .
\end{eqnarray}
Given that the pressure of mimetic matter reads $p = -V$, we immediately see that the EoS for mimetic matter is $w = \tilde{w}$, demonstrating that mimetic matter, when subdominant, can imitate the EoS of the dominant energy component \cite{m2}. A comment is in order here. Mimetic matter can only be subdominant if $\alpha/w \ll 1$. If this condition is not satisfied, mimetic matter will only start imitating the dominant matter component at late times, while acting as a cosmological constant at earlier times.

\subsubsection{Power law potential}

We can consider an arbitrary power law potential:
\begin{eqnarray}
V(\phi) = \alpha \phi ^n = \alpha t^n \, ,
\end{eqnarray}
for which the solution of Eq.(\ref{yeq}) can be written in terms of the Bessel functions \cite{m2}:
\begin{eqnarray}
y = t^{\frac{1}{2}}Z_{\frac{1}{n+2}}\left ( \frac{\sqrt{-3\alpha}}{n+2}t^{\frac{n+2}{2}} \right ) \, .
\end{eqnarray}
For $n < -2$ (with $n = -2$ corresponding to the case we have studied previously) the limiting behaviour of the scale factor is that of a dust-dominated Universe, with EoS $w = 0$. For $n > -2$ and $\alpha < 0$ (which corresponds to a positive pressure), the corresponding solution is a singular oscillating Universe. For $n > -2$ and $\alpha > 0$ instead, the pressure is negative and hence we expect accelerating solutions \cite{m2}. In fact, $n = 0$ corresponds to a cosmological constant as expected (the potential is simply a constant), whereas $n = 2$ gives an inflationary expansion solution with scale factor:
\begin{eqnarray}
a \propto t^{-\frac{1}{3}}e^{\sqrt{\frac{\alpha}{12}}t^2} \, ,
\end{eqnarray}
which resembles that of chaotic inflation sourced by a quadratic potential \cite{m2}.

\subsubsection{Inflation in mimetic gravity}

One can always reconstruct the appropriate potential for the mimetic field which can provide an inflationary solution. The method is very simple: choose a desired expansion history of the Universe [encoded in the Hubble parameter $H$ or equivalently in the scale factor $a(t)$], find the corresponding $y$-parameter through $a = y^{\frac{2}{3}}$, then invert Eq.(\ref{yeq}) to find the corresponding potential which can provide the desired expansion:
\begin{eqnarray}
V(\phi) = V(t) = \frac{4\ddot{y}}{3y} \, .
\end{eqnarray}
As an example, we can consider the following potential \cite{m2}:
\begin{eqnarray}
V(\phi) = \frac{\alpha \phi ^2}{e^{\phi}+1} \, ,
\end{eqnarray}
whose corresponding solution for the scale factor is exponential ($a \propto e^{-t^2}$) for large negative times, and corresponding to EoS $w = 0$ (i.e. dust, $a \propto t^{2/3}$) at late times. Thus we see that with this potential mimetic gravity can provide us with an inflationary solution with graceful exit to the matter dominated era \cite{m2}. \\

Another interesting possibility is given by an exponential potential \cite{saadi}:
\begin{eqnarray}
V(\phi) = \alpha e^{-\kappa\phi} = \alpha e^{-\kappa t} \, .
\end{eqnarray}
In this case, the scale factor can be expressed in terms of Bessel functions of the first and second kind \cite{saadi}:
\begin{eqnarray}
a(t) = \left [ \beta J_0 \left ( \frac{\sqrt{-3\alpha}}{\kappa}e^{-\frac{kt}{2}} \right ) + \gamma Y_0 \left ( \frac{\sqrt{-3\alpha}}{\kappa}e^{-\frac{kt}{2}} \right ) \right ] ^{\frac{2}{3}} \, ,
\end{eqnarray}
where $\beta$ and $\gamma$ are integration constants, and $J_0$ and $Y_0$ are the modified Bessel functions of order zero, of the first and second kind respectively. At late times, the behaviour of the scale factor is that of a matter-dominated Universe, i.e. with EoS $w = 0$ ($a \propto t^{2/3}$). At early times the behaviour of the scale factor depends on the sign of $\alpha$, in particular providing us with an inflationary solution for $\alpha > 0$, whereas the solution is a bouncing non singular Universe for $\alpha < 0$. A similar behaviour can be obtained if one chooses the potential:
\begin{eqnarray}
V(\phi) = \frac{\alpha \phi^{2n}}{e^{\kappa \phi} + 1} = \frac{\alpha t^{2n}}{e^{\kappa t} + 1} \, ,
\end{eqnarray}
that is, a solution with inflation at early times and matter domination at late times. More complicated potentials which can reproduce qualitatively similar behaviours were studied in \cite{khalifeh}.

\subsubsection{Bouncing Universes in mimetic gravity}

As we have already seen in previous cases, one can easily construct bouncing solutions in mimetic gravity. Let us work through one further example here. Consider a potential of the form:
\begin{eqnarray}
V(\phi) = \frac{4}{3}\frac{1}{(1+\phi^2)^2} = \frac{4}{3}\frac{1}{(1+t^2)^2} \, .
\end{eqnarray}
As usual, the scale factor can be determined by solving Eq.(\ref{yeq}), which yields \cite{m2}:
\begin{eqnarray}
a(t) = \left [ \sqrt{t^2+1} \left ( 1 + \beta \arctan t \right ) \right ] ^{\frac{2}{3}} \, .
\end{eqnarray}
If we set the integration constant $\beta$ to 0, the corresponding energy density and pressure (one again, refer to Eq.(\ref{energy}) and recall that $p = -V$) are given by:
\begin{eqnarray}
\rho = \frac{4}{3}\frac{t^2}{(1+t^2)^2} \quad , \quad p = -\frac{4}{3}\frac{1}{(1+t^2)^2} \, .
\end{eqnarray}
At very early times (large negative $t$) the EoS approaches $w \rightarrow 0$, the Universe is dominated by dust and contracts. At a certain time corresponding to $|t| \sim 1$, the energy density drops suddenly to zero, after which the Universe begins expanding. During the first instants of the expansion (within one Planckian time), the energy density of the Universe is Planckian, but subsequently drops as the expansion proceeds as a conventional expansion in a dust dominated Universe. The interesting feature of this potential is that the EoS crosses the phantom divide without singularity. This remains true even in the general case where the integration constant $\beta$ is non-zero, provided $|\beta| < 2/\pi$ \cite{m2}. \\

In the case we have just examined, the bounce occurs at the Planck scale, and hence the classical analysis we provided might not be valid as quantum gravity effects would be playing an important role. However, a minimal modification allows to lower the scale of the bounce, and correspondingly increase the duration of the bounce (which now lasts more than a Planckian time). The corresponding potential which can provide this behaviour is given by \cite{m2}:
\begin{eqnarray}
V(\phi) = \frac{4}{3}\frac{\alpha}{(\phi_0^2 + \phi^2)^2} = \frac{4}{3}\frac{\alpha}{(t_0^2 + t^2)^2} \, .
\end{eqnarray}
Although we won't show the solution explicitly, in this case the scale of the bounce is reduced to $\alpha/t_0^2$ and the duration of the bounce is now $t_0$ \cite{m2}.

\subsection{Mimetic $\mathbf{F(R)}$ gravity}

The next step, which was performed by Nojiri and Odintsov is to generalize mimetic gravity to mimetic $F(R)$ gravity \cite{nompla}. In this theory we expect to have two additional degrees of freedom compared to GR: the constrained (non-propagating) scalar degree of freedom of mimetic gravity, and the additional scalar degree of freedom arising from the $F(R)$ term. The action of the theory is given by \cite{nompla}:
\begin{eqnarray}
I = \int d^4x \sqrt{-g(\tilde{g}_{\mu \nu},\phi)} \ \left [ F(R(\tilde{g}_{\mu \nu},\phi)) + {\cal L}_m \right ] \, ,
\label{mimeticfr}
\end{eqnarray}
where as usual the relation between the physical and auxiliary metric, and the mimetic field, is given by Eq.(\ref{mimeticmetric}), and the constraint equation Eq.(\ref{norm}) has to be satisfied. Because of this, the action of mimetic $F(R)$ gravity can be equally written employing a Lagrange multiplier field analogously to Eq.(\ref{actionlm}) \cite{nompla}:
\begin{eqnarray}
I = \int d^4x \sqrt{-g} \ \left [ F(R(g_{\mu \nu})) - V(\phi) + \lambda (g^{\mu \nu}\partial_{\mu}\phi\partial_{\nu}\phi + 1) + {\cal L}_m \right ] \, ,
\label{mimeticfrp}
\end{eqnarray}
where the Lagrange multiplier enforces the constraint on the gradient of the mimetic field, and in addition we have added a potential for the mimetic field. \\

The equations of motion of the theory are slightly more complicated than that of conventional mimetic gravity. Varying with respect to the metric gives the gravitational field equations \cite{nompla}:
\begin{eqnarray}
0 &=& \frac{1}{2}g_{\mu \nu}F(R) - R_{\mu \nu}F'(R) + \nabla_{\mu}\nabla_{\nu}F'(R) - g_{\mu \nu}\Box F'(R) \nonumber \\
&+& \frac{1}{2}g_{\mu \nu} \left [ -V(\phi) + \lambda (g^{\alpha \beta}\partial_{\alpha}\phi\partial_{\beta}\phi) \right ] - \lambda \partial_{\mu}\phi\partial_{\nu}\phi + \frac{1}{2}T_{\mu \nu} \, .
\label{eq1}
\end{eqnarray}
Variation with respect to the mimetic field instead yields the following equation \cite{nompla}:
\begin{eqnarray}
2\nabla ^{\mu}(\lambda \partial_{\mu}\phi) + \frac{dV}{d\phi} = 0 \, .
\label{eq2}
\end{eqnarray}
As usual, by construction, variation with respect to the Lagrange multiplier gives the mimetic constraint:
\begin{eqnarray}
g^{\mu \nu}\partial_{\mu}\phi\partial_{\nu}\phi = -1 \, .
\label{eq3}
\end{eqnarray}

As we have mentioned previously, in mimetic $F(R)$ gravity one has two additional degrees of freedom. Therefore, by appropriately tuning either or both the potential for the mimetic field, or the form of the $F(R)$ function, one can reconstruct basically any desired expansion history of the Universe. On the other hand, the interpretation of the cosmological role played by mimetic dark matter remains the same as in mimetic gravity \cite{nompla}. \\

Let us proceed to study some of the properties of mimetic $F(R)$ gravity in a cosmological setting. As usual, we consider a flat FLRW Universe, and we model the matter contribution as that of a perfect fluid with energy density $\rho$ and pressure $p$. Assuming that the mimetic field depends only on time, Eqs.(\ref{eq1},\ref{eq2},\ref{eq3}) can be expressed as follows \cite{nompla}:
\begin{eqnarray}
0 &=& -F(R) + 6(\dot{H} + H^2)F'(R) - 6H\frac{dF'(R)}{dt} - \lambda (\dot{\phi}^2 + 1) + V(\phi) + \rho \, , \\
\label{first}
0 &=& F(R) - 2(\dot{H} + 3H) + 2\frac{d^2F'(R)}{dt^2} + 4H\frac{dF'(R)}{dt} -\lambda (\dot{\phi}^2 - 1) - V(\phi) + p \, , \\
\label{second}
0 &=& 2\frac{d}{dt}(\lambda \dot{\phi}) + 6H\lambda \dot{\phi} -\frac{dV}{d\phi} \, , \\
\label{third}
0 &=& \dot{\phi}^2 - 1 \, .
\label{fourth}
\end{eqnarray}
The last equation shows that up to an integration constant, which we can set to zero, the mimetic field can be identified with time just as in ordinary mimetic gravity. Thus, Eq.(\ref{second}) can be expressed as follows:
\begin{eqnarray}
0 = F(R) - 2(\dot{H} + 3H) + 2\frac{d^2F'(R)}{dt^2} + 4H\frac{dF'(R)}{dt} - V(t) + p \, ,
\end{eqnarray}
which, if we assume that the contribution of ordinary matter is negligible ($\rho = p = 0$), reduces to:
\begin{eqnarray}
V(t) = F(R) - 2(\dot{H} + 3H) + 2\frac{d^2F'(R)}{dt^2} + 4H\frac{dF'(R)}{dt} \, . 
\label{reconstruct}
\end{eqnarray}
On the other hand, Eq.(\ref{first}) can be solved for $\lambda$ as follows:
\begin{eqnarray}
\lambda (t) =  -\frac{1}{2}F(R) + 3(\dot{H} + H^2)F'(R) - 3H\frac{dF'(R)}{dt} \, ,
\end{eqnarray}
which shows that Eq.(\ref{fourth}) is automatically satisfied. \\

The above equations put on a quantitative footing the statement we previously made: namely, that by tuning the behaviour of either or both the two additional scalar degrees of freedom, we can reconstruct any possible expansion history of the Universe \cite{nompla}. For instance, one can imagine fixing the form of the scalar potential and then reconstruct the form of $F(R)$ which gives the wanted evolution [encoded in $H(t)$ or, equivalently, $a(t)$]. Alternatively, one can start from a given form of $F(R)$ which might not admit the wanted evolution (e.g. matter dominated-like expansion followed by accelerated expansion) and reconstruct the form of the scalar potential which can allow for such expansion. It should also be remarked that any solution in conventional $F(R)$ gravity is also a solution in mimetic $F(R)$ gravity, but the converse is not true. In \cite{nompla} specific solutions which allow unification of early-time inflation and late-time acceleration with intermediate matter domination era, as well as bouncing Universes, are studied and it is shown that they can be implemented in mimetic $F(R)$ gravity. Of course, the exact forms of the mimetic potential or the $F(R)$ function in these cases are quite complicated, nonetheless the study serves as a proof of principle that, in such theories, one can realize any given expansion history of the Universe without the need for dark components, which remains the main goal of modified theories of gravity. \\

Three further recent studies by Odintsov and Oikonomou \cite{oikonomou,oikonomouaccelerating,oikonomouapss} have demonstrated how one can, in mimetic $F(R)$ gravity, realize inflationary cosmologies which are compatible with Planck and BICEP2/Keck Array constraints on the scalar spectral index and on the tensor-to-scalar ratio, $n_s$ and $r$, respectively. Both the reconstruction (determination of the potential once the form of $F(R)$ and the evolutionary history of the Universe are given) and the inverse reconstruction (determination of the form of $F(R)$ once the potential and the evolutionary history of the Universe are given) are studied in detail and it is demonstrated that several viable options for realizing inflation compatibly with observational constraints are possible. However, the studies also point out a possible weakness of mimetic $F(R)$ gravity in this respect: namely, the forms of both the mimetic potential and the $F(R)$ function can become extremely complicated. The forms of both the mimetic potential and the Lagrange multiplier increase in complexity as the complexity of the $F(R)$ form increases. Therefore, these and other studies on mimetic $F(R)$ gravity and extensions thereof should not be viewed as the ultimate cosmological theory of everything, but rather as a proof of principle that within these theories one can reproduce basically any cosmological scenario and thus solve the ``dark Universe'' problems, although this might come at the cost of sacrificing simplicity. \\

In addition, \cite{oikonomou,oikonomouaccelerating,oikonomouapss} also remarked that, although in principle the forms of the mimetic potential and the $F(R)$ function can be arbitrary, it must be kept in mind that in order to realize viable inflation, a mechanism for graceful exit to the conventional radiation dominated era must be achieved. This entails ensuring that the theory contain an unstable de Sitter vacua, which eventually becomes the cosmological attractor of the dynamical system. It is precisely the functional form of $F(R)$ which has to ensure that graceful exit takes place. Therefore, in mimetic gravity, although in principle the form of the potential is arbitrary, the same cannot be said about the functional form of $F(R)$, which has to be such as to ensure graceful exit from inflation. Therefore, in the interest of simplicity, a practical approach to constructing a minimal model of mimetic $F(R)$ gravity with the desired inflationary properties would be to choose the simplest possible functional form of $F(R)$ which ensures graceful exit from inflation, then perform the reconstruction technique to determine the form of the mimetic potential which allows the desired expansion history following inflation to be realized. Another possible solution, which we will discuss shortly, is to consider $F(R,\phi)$ inflation \cite{myrzakulov,myrzakul}, where a dynamical scalar field $\phi$ is coupled to gravity. \\

\subsubsection{Late-time evolution in mimetic gravity}

So far we have discussed mimetic gravity and variants thereof at early times. That is, at the epoch when primordial curvature perturbations were generated. However, it is also interesting to consider late-time evolution in mimetic gravity. The equations of motion are incredibly complex and in principle do not allow for analytical solutions. However, this complexity can be bypassed by means of the method of dynamical analysis (see e.g. \cite{dynamical1,dynamical2,dynamical3,dynamical4}), which gives information about the global behaviour of solutions. In particular, one proceeds by transforming the equations of motion into their autonomous form and extract the critical points. Subsequently perturbations are linearized around these critical points and expressed in terms of the perturbation matrix, the eigenvalues of which determine the type and stability of the critical points. \\

A detailed dynamical analysis of mimetic $F(R)$ gravity was presented in \cite{leon}. This type of analysis allows to bypass the complexity of the equations of motion by extracting critical points and studying the corresponding observables, such as the energy densities of the various energy components, the corresponding EoS, and the deceleration parameter. In particular, the analysis finds that the only stable critical points, i.e. those that can play the role of attractors at late times, are those that exist in $F(R)$ gravity as well. In other words, stable solutions in mimetic $F(R)$ gravity can only affect the expansion history of the Universe at early and intermediate times, whereas at late times the expansion history has to coincide with that driven by conventional $F(R)$ gravity. An immediate implication of this finding is that, although mimetic $F(R)$ gravity could drive inflation differently from $F(R)$ gravity, the late-time acceleration of the Universe in these theories has to coincide with the usual $F(R)$ gravity one \cite{leon}. However, these conclusions have been reached only by studying the theory at the level of the background. It is expected that different conclusions would be reached if the same study would be performed at the level of perturbations. This is true because the new terms present in the equations of motion of mimetic $F(R)$ gravity compared to conventional $F(R)$ gravity, can contribute to the perturbation equations, although they do not contribute at the background level \cite{leon}. Finally, the energy conditions required to avoid the Dolgov-Kawasaki instability in mimetic gravity were studied in \cite{shiravand}, which found that these are the same as in conventional $F(R)$ gravity. \\

As we mentioned above, the conclusions reached about the late-time evolution in mimetic $F(R)$ gravity hold only at the background level. However, in conventional $F(R)$ gravity, there exists a serious problem during the late-time evolution at the perturbation level. Namely, that of dark energy oscillations \cite{oscillations1} (see also e.g. \cite{oscillations2,oscillations3,oscillations4,oscillations5}). The degree of freedom associated to the modification of GR (that is, $dF(R)/dR$) leads to high frequency oscillations of the dark energy around the line of the phantom divide during matter era. As a consequence, some derivatives of the Hubble parameter may diverge and become singular, and the solution is unphysical. Usually in conventional $F(R)$ gravity the problem is solved by adding power-law modifications by hand. \\

In \cite{odintsovoscillations}, it was instead argued that in mimetic $F(R)$ gravity it is possible to overcome the problem by a suitable choice of the potential. By appropriately choosing the potential and the Lagrange multiplier, it is possible to damp the oscillations within a mimetic $F(R)$ model whose corresponding conventional $F(R)$ model suffered from the oscillations problems. The oscillations die out for redshifts $z \leq 3$, so there is no issue with dark energy oscillations at our current epoch. Moreover, the values of the dark energy equation of state and the total equation of state are very close to the observed values. The model can in principle be discerned from $F(R)$ gravity in that the predicted growth factor is lower in magnitude, a very testable prediction in view of future experiments, further supporting the viability of mimetic $F(R)$ gravity as a cosmological framework.

\subsubsection{Mimetic $\mathbf{F(R,\phi)}$ gravity}

The question of constructing a theoretically motivated but at the same time simple model for mimetic $F(R)$ gravity which provides an early-time inflation epoch, but at the same time graceful exit from the latter, was addressed in \cite{myrzakulov}. Here, a model of mimetic $F(R,\phi)$ gravity was considered, where $\phi$ is a scalar field coupled to gravity.  We will not go into the details of the work, for which we refer the reader to the original paper \cite{myrzakulov}. The basic idea is to use the $F(R)$ sector to reproduce a variety of cosmological scenarios: among the ones considered in the paper were accelerated cosmologies at high and low curvatures (thus unifying inflation and late-time acceleration), with Einstein gravity at intermediate curvatures. In particular, the accelerated cosmologies are realized by making use of a ``switching-on'' cosmological constant. The dynamical field $\phi$ evolves in such a way to allow for graceful exit from the inflationary period, thus making the vacua of the first de Sitter period (corresponding to inflation) unstable. Entry into the late-time accelerated epoch, represented by a stable de Sitter attractor, is also made possible by the dynamical field, which thus links all epochs of the expansion history of the Universe in an unified way. In the minimal case studied in \cite{myrzakulov}, the mimetic component ensures the presence of cosmological non-baryonic dark matter although, as we have extensively discussed, it is possible by adding a suitable potential for the mimetic field to obtain similar solutions, but with a different form of $F(R)$.

\subsubsection{Non-local mimetic $\mathbf{F(R)}$ gravity}

A further extension of mimetic $F(R)$ gravity was presented in \cite{sebastianinonlocal}, which embeds the theory into the framework of non-local theories of gravity. Recall these theories were first presented in \cite{deserwoodard}, inspired by quantum loop corrections. These theories feature non-local operators (i.e. inverse of differential operators) of the curvature invariants. The prototype of non-local mimetic $F(R)$ gravity is given by the action \cite{sebastianinonlocal}:
\begin{eqnarray}
I = \int d^4x\sqrt{-g} \ \left [ R (1 + f(\Box ^{-1}R)) + \lambda (g^{\mu \nu}\partial_{\mu}\phi\partial_{\nu}\phi + 1) - V(\phi) \right ] \, ,
\label{nl}
\end{eqnarray}
where as usual $\phi$ is the mimetic field, and the Lagrange multiplier term enforces the constraint on its gradient. It is actually more useful to introduce an additional scalar field $\psi$, which allows to translate the action given by Eq.(\ref{nl}) to a local scalar-tensor form, as follows \cite{sebastianinonlocal}:
\begin{eqnarray}
I = \int d^4x\sqrt{-g} \ \left [ R(1 + f(\psi)) + \xi (\Box \psi - R) + \lambda (g^{\mu \nu}\partial_{\mu}\phi\partial_{\nu}\phi + 1) - V(\phi) \right ] \, ,
\end{eqnarray}
where $\xi$ is an additional Lagrange multiplier which enforces the constraint on the scalar field $\psi$:
\begin{eqnarray}
\Box \psi = R \, .
\end{eqnarray}
Aside from the two constraint equations obtained by varying the action with respect to the Lagrange multipliers, variation of the action with respect to the yields the gravitational field equations \cite{sebastianinonlocal}:
\begin{eqnarray}
R_{\mu \nu}(1 + f(\psi) - \xi) &-& \frac{1}{2}g_{\mu \nu} R(1 + f(\psi) - \xi) - \partial_{\rho}\xi\partial^{\rho}\psi = \frac{1}{2}(\partial_{\mu}\xi\partial_{\nu}\psi + \partial_{\mu}\psi\partial_{\nu}\xi) \nonumber \\
&-&(g_{\mu \nu}\Box - \nabla_{\mu}\nabla_{\nu})(f(\psi) - \xi) - \lambda\partial_{\mu}\phi\partial_{\nu}\phi - g_{\mu \nu}V(\phi) \, ,
\end{eqnarray}
Instead, variation with respect to the two scalar fields leads to the following equations of motion \cite{sebastianinonlocal}:
\begin{eqnarray}
\Box \xi + f'(\psi)R &=& 0 \, , \\
-\frac{1}{\sqrt{-g}}\partial_{\nu}(\sqrt{-g}\partial^{\nu}\phi) &=& \frac{1}{2}\frac{dV}{d\phi} \, .
\end{eqnarray}
In \cite{sebastianinonlocal}, this model was studied in detail making use of the reconstruction technique. In particular, two forms for the $f(\psi)$ function have been studied: exponential and power-law. It was shown that appropriate choices for the mimetic potential, as expected, can give the desired expansion history of the Universe, which in the cases studied included viable inflation unified with late-time acceleration with intermediate epoch compatible with Einstein's gravity and cosmological dark matter provided by the mimetic field, as well as solutions with cosmological bounces.

\subsection{Unimodular mimetic gravity}

As we have seen, mimetic gravity provides a geometric explanation for dark matter in the Universe, with dark matter emerging as an integration constant as a result of gauging local Weyl invariance, without the need for additional fluids. An older theory, known as unimodular gravity \cite{unimodular1} (see also \cite{bufalo,unimodular2,unimodular3,unimodular4,unimodular5,unimodular6,unimodular7,unimodular8,unimodular9,
unimodular10,unimodular11,unimodular12,unimodular13,unimodular14,unimodular15,unimodular16,unimodular17,unimodular18,
unimodular19,unimodular20}), had instead been proposed much earlier to solve, in a geometrical fashion as well, another of the conundrums of modern cosmology: the dark energy problem. In this framework, dark energy emerges in the form of a cosmological constant from the trace-free part of Einstein's field equations, with the trace-free part which results in turn by enforcing the condition that the square root of (minus) the determinant of the metric is equal to 1, or in general a constant. It would therefore be interesting to combine the two different approaches of mimetic gravity and unimodular gravity into a single framework which could geometrically explain both dark matter and dark energy by a vacuum theory, without need for additional fluids. This is the proposal of Nojiri and Odintsov in \cite{odintsovunimodular}. \\

In order to combine mimetic gravity and unimodular gravity it is necessary to enforce two constraints. The first is the constraint on the gradient of the mimetic field Eq.(\ref{norm}), whereas the second is the unimodular constraint:
\begin{eqnarray}
\sqrt{-g} = 1 \, .
\label{unimodular}
\end{eqnarray}
In order to enforce these two constraints, it is conceptually simple to make use of two Lagrange multipliers. This approach has two advantages: first, it keeps the two concepts of mimetic and unimodular gravity separate, and facilitates the extraction of physical information. Second, it is as usual more convenient to have the two constraints emerge from the equations of motion. The action for unimodular mimetic gravity thus reads \cite{odintsovunimodular}:
\begin{eqnarray}
I = \int d^4x \ \left [ \sqrt{-g} (R - V(\phi) - \eta (g^{\mu \nu}\partial_{\mu}\phi\partial_{\nu}\phi + 1) - \lambda ) + \lambda \right ] \, .
\label{unimodularaction}
\end{eqnarray}
where variation with respect to the Lagrange multiplier $\eta$ enforces the mimetic constraint Eq.(\ref{norm}), whereas variation with respect to the Lagrange multiplier $\lambda$ enforces the unimodular constraint Eq.(\ref{unimodular}). \\

The equations of motion for the gravitational field are obtained by varying the action with respect to the metric, and are given by \cite{odintsovunimodular}:
\begin{eqnarray}
0 = \frac{1}{2}g_{\mu \nu} \left ( R - V(\phi) - \eta(g^{\alpha \beta}\partial_{\alpha}\phi\partial_{\beta}\phi + 1) - \lambda \right ) - R_{\mu \nu} + \eta \partial_{\mu}\phi\partial_{\nu}\phi + \frac{1}{2}T_{\mu \nu} \, ,
\end{eqnarray}
whereas variation with respect to the mimetic field yields the usual equation of motion:
\begin{eqnarray}
0 = 2\nabla ^{\mu}(\lambda\partial_{\mu}\phi) - \frac{dV}{d\phi} \, .
\end{eqnarray}
Although we do not show the steps explicitly, for which we refer the reader to the original paper \cite{odintsovunimodular}, in the usual FLRW setting it is possible to manipulate the Einstein equations in order to get the following reconstruction equation for the mimetic potential $V(\phi) = V(t)$ (as usual, the mimetic field can be identified with time):
\begin{eqnarray}
&&V(\phi) = V(t) =\nonumber\\&&\hspace{-0.5cm} \frac{a(t)^2}{3}\int_0^{\phi} dt \ a(t)^{-2} \left [ -6H(t)p(t) - 2\frac{dp(t)}{dt} + 2 \left ( -18H(t)^3 - 6H(t)\frac{dH(t)}{dt} + 4\frac{d^2H(t)}{dt^2} \right ) \right ] \, .
\label{unimimpotential}
\end{eqnarray}
The content of the above equation is clear: as in all extensions so far discussed of mimetic gravity, one can always reconstruct the potential for the mimetic field which can provide the desired expansion encoded in $H(t)$ or $a(t)$. The reconstruction technique is very powerful although, as we have seen, the corresponding potentials are complicated and somewhat hard to justify from first principles, although the reconstruction technique serves in this case as a proof of principle tool. \\

Having made this consideration, let us consider a few examples where the reconstruction technique is applied. Let us consider the following simple potential \cite{odintsovunimodular}:
\begin{eqnarray}
V(\phi) = 12e^{2H_0\phi}H_0^3\phi \, .
\end{eqnarray}
It can be easily shown that it leads to the following solution for the scale factor $a(t)$ and the Hubble parameter $H(t)$:
\begin{eqnarray}
a(t) = e^{H_0t} \quad , \quad H(t) = H_0
\end{eqnarray}
i.e. a de Sitter cosmology. The functional forms of the Lagrange multiplier are also quite simple, and are given by the following:
\begin{eqnarray}
\lambda (t) = 6H_0^2(1 + 2e^{2H_0t}H_0t) \quad , \quad \eta (t) = 3H_0^2 \, .
\end{eqnarray}
Another choice for the potential which leads to a physically interesting solution is the following \cite{odintsovunimodular}:
\begin{eqnarray}
V(\phi) =  -\frac{8\phi^{-2 + \frac{4}{3(1 + w)}}(1 + 5w + 2w^2)}{9(1 + w)^3} \, ,
\end{eqnarray}
for which the scale factor and the Hubble parameter read:
\begin{eqnarray}
a(t) = t^{\frac{2}{3(1+w)}} \quad , \quad H(t) = \frac{2}{3t(1 + w)} \, .
\end{eqnarray}
The above solution is that corresponding to an Universe dominated by a fluid with EoS $w$. The two Lagrange multipliers are given by:
\begin{eqnarray}
\lambda (t) = \frac{8(-3w(1+w) + t^{\frac{4}{3(1+w)}}(1 + 5w + 2w^2)}{9t^2(1 + w)^3} \quad , \quad \eta (t) = \frac{4(2+w)}{3t^2(1 + w)^2} \, .
\end{eqnarray}
Thus we see that with two relatively simple choices of potential it is possible to reconstruct two important expansion histories of the Universe: the late-time de Sitter phase and the expansion dominated by a perfect fluid with arbitrary EoS. Although we will not discuss this case explicitly here, for which instead we redirect the reader to \cite{odintsovunimodular}, it is possible by a choice of a more complicated potential, to realize a viable inflationary model within unimodular mimetic gravity, which is compatible with bounds from Planck and BICEP2/Keck Array. Moreover, it has been shown that graceful exit from these type of inflationary periods can be achieved by ensuring that the corresponding de Sitter vacua which drives the period of accelerated expansion is unstable. \\

Two further comments are in order here. First, it is possible to provide an effective fluid description of unimodular mimetic gravity \cite{odintsovunimodular}. Namely, manipulation of the Einstein equations shows that the contribution of the unimodular and mimetic parts of the action can be interpreted as that of a perfect fluid carrying energy density $\rho$ and pressure $p$ as follows:
\begin{eqnarray}
\rho = G - T - 4\tilde{V} \quad , \quad p = -\tilde{V} \, ,
\end{eqnarray}
where $\tilde{V}$ is defined as:
\begin{eqnarray}
\tilde{V} = -\lambda (t) - V(t) \, .
\end{eqnarray}
Furthermore, the effective energy density and pressure defined as per above satisfy the continuity equation. \\

The second comment is related to the fact that the unimodular constraint is enforced in a non-covariant way (c.f. the action given by Eq.(\ref{unimodularaction}), where one of the terms in $\lambda$ is not multiplied by $\sqrt{-g}$). It is nonetheless possible to present a covariant formulation of unimodular mimetic gravity via the following action:
\begin{eqnarray}
I = \int d^4x \ \left [ \sqrt{-g} \left ( R - \eta (g^{\mu \nu}\partial_{\mu}\phi\partial_{\nu}\phi + 1) - \lambda \right ) + \lambda \epsilon^{\mu \nu \rho \sigma}\partial_{\mu}a_{\nu \rho \sigma} \right ] \, ,
\end{eqnarray}
where $a_{\nu \rho \sigma}$ is a three-form field, variation of which gives the constraint $\partial_{\mu}\lambda = 0$, implying that the Lagrange multiplier $\lambda$ is constant. On the other hand, the covariant version of the unimodular constraint is obtained by variation with respect to $\lambda$, from which one is left with:
\begin{eqnarray}
\sqrt{-g} = \epsilon^{\mu\nu\rho\sigma}\partial_{\mu}a_{\nu\rho\sigma} \, .
\end{eqnarray}
Further manipulation, for which we refer the reader to \cite{odintsovunimodular}, shows that the Friedmann equations one obtains from the covariant version of unimodular mimetic gravity are equivalent to those of the non-covariant version, and thus one can reproduce precisely the same cosmological scenarios in both theories.

\subsubsection{Unimodular mimetic $F(R)$ gravity}

A minimal extension of the unimodular mimetic gravity framework we have discussed so far is to consider unimodular mimetic $F(R)$ gravity, which is described by the action \cite{odintsovum}:
\begin{eqnarray}
I = \int d^4x \ \left [ \sqrt{-g} (F(R) - V(\phi) - \eta (g^{\mu \nu}\partial_{\mu}\phi\partial_{\nu}\phi + 1) - \lambda ) + \lambda \right ] \, .
\end{eqnarray}
As expected, the equations of motion are slightly more complicated than in the unimodular mimetic case, but no conceptual difficulty is added. Specifically, variation with respect to the two Lagrange multipliers enforces the usual unimodular and mimetic constraints, whereas variation with respect to the metric gives rise to the equations for the gravitational field \cite{odintsovum}:
\begin{eqnarray}
\frac{g_{\mu \nu}}{2}(F(R) - V(\phi) + \eta(g^{\alpha \beta}\partial_{\alpha}\phi\partial_{\beta}\phi + 1) - \lambda) - R_{\mu \nu}F'(R) - \eta\partial_{\mu}\phi\partial_{\nu}\phi + \nabla_{\mu}\nabla_{\nu}F'(R) - g_{\mu \nu}\Box F'(R) = 0 \, .
\end{eqnarray}
Finally, variation with respect to the mimetic field yields the following equation:
\begin{eqnarray}
0 = 2\nabla ^{\mu}(\lambda\partial_{\mu}\phi) - \frac{dV}{d\phi} \, .
\end{eqnarray}
The more complex structure of the equations of motion complicates the reconstruction procedure, i.e. the equivalent of Eq.(\ref{unimimpotential}), which now reads \cite{odintsovum}:
\begin{eqnarray}
V(t) = \frac{a(t)^\frac{3}{2}}{2}\int dt \ a(t)^{-\frac{3}{2}}f(t) \, ,
\end{eqnarray}
where the function $f(t)$ is given by:
\begin{eqnarray}
f(t) &=& - [ 18H(t) (\dot{H} + H^2)F'(R) - 6H^2\frac{dF'(R)}{dt} - 18\dot{H}H^3 + 6H\frac{d^3F'(R)}{dt^3} + (6\ddot{H} + 2\dot{H}H)F'(R) \nonumber \\
&+& H\dot{H}\frac{dF'(R)}{dt} + 6H^2\frac{dF'(R)}{dt} - 2H\frac{d^2F'(R)}{dt^2} - 2(\ddot{H} + 6\dot{H}H) + 2\frac{d^2F'(R)}{dt^2} ]
\end{eqnarray}
Despite the increased complexity of the equations of motion, the same considerations apply as for unimodular mimetic gravity, as well as all extensions of mimetic gravity hereto considered. Namely, it is always possible to reconstruct any viable cosmological expansion scenario, including unification of inflation and late-time acceleration with intermediate radiation and matter domination, with graceful exit from inflation triggered by an unstable de Sitter vacua. This can be achieved by appropriately choosing either or both the form of the mimetic potential or the function $F(R)$. The price to pay would eventually be a considerable complexity in the functional form of both, which of course does not represent a first principle obstacle \cite{odintsovum}.

\subsection{Mimetic Horndeski gravity}

One can further consider more general scalar-tensor theories, which can be ``mimetize'' according to the procedures we have describes so far, namely through a singular disformal transformation or through a Lagrange multiplier term in the action enforcing the mimetic constraint. In fact, analogously to GR, one can show that the most general scalar-tensor model is invariant under disformal transformations, provided the latter is invertible. This has been shown in all generality in \cite{arroja}. One can then ``mimetize'' such theories by considering the following action \cite{arroja}:
\begin{eqnarray}
I = \int d^4x \sqrt{-g} \ \left [ {\cal L}(g_{\mu \nu}, \partial_{\lambda_1}g_{\mu \nu},...,\partial_{\lambda_1}...\partial_{\lambda_p}g_{\mu \nu}, \phi, \partial_{\lambda_1}\phi,...,\partial_{\lambda_1}...\partial_{\lambda_q}\phi) + \lambda(g^{\mu\nu}\partial_{\mu}\phi\partial_{\nu}\phi + 1) \right ] \, ,
\end{eqnarray}
where $p, q \geq 2$ are integers and ${\cal L}$ is the Lagrangian density which is a function of the metric and the mimetic field. In general, the constraint enforced by the Lagrange multiplier can be generalized to $b(\phi)g^{\mu\nu}\partial_{\mu}\phi\partial_{\nu}\phi = -1$, but for the sake of simplicity we will set $b(\phi) = 1$ here, and redirect the reader to the work of \cite{arroja} for more general discussions, and for the explicit form of the equations of motion. In fact, setting $b(\phi) \neq 1$ is basically equivalent to assigning a potential to the mimetic field. In \cite{arroja} it was shown that the two approaches to mimetic Horndeski gravity, namely singular disformal transformation and Lagrange multiplier, are equivalent. \\

Of course, the considerations made above can be applied in the case of a specific scalar-tensor model, namely Horndeski gravity \cite{horndeski} (see for instance \cite{horndeski1,horndeski2,horndeski3,horndeski4,horndeski5,horndeski6,horndeski7,horndeski8,horndeski9,horndeski10,
horndeski11,horndeski12,horndeski13,horndeski14,horndeski15,horndeski16,horndeski17,horndeski18,horndeski19} for further recent work on the topic of Horndeski gravity and theories beyond Horndeski). Recall that Horndeski gravity is the most general 4D local scalar-tensor theory with equations of motion no higher than second order. The Horndeski action can be written as a sum of four terms:
\begin{eqnarray}
I = \int d^4x \sqrt{-g} \ {\cal L}_H = \int d^4x \sqrt{-g} \ \sum _n {\cal L}_n \, ,
\end{eqnarray}
where the ${\cal L}_n$s read:
\begin{eqnarray}
{\cal L}_0 &=& K(X,\phi) \, , \\
{\cal L}_1 &=& -G_3(X,\phi)\Box\phi \, , \\
{\cal L}_2 &=& G_{4,X}(X,\phi) \left [ (\Box \phi)^2 - (\nabla_{\mu}\nabla_{\nu}\phi)^2 \right ] + RG_4(X,\phi) \, , \\
{\cal L}_3 &=& -\frac{1}{6}G_{5,X}(X,\phi) \left [ (\Box \phi)^3 - 3\Box \phi (\nabla_{\mu}\nabla_{\nu}\phi)^2 + 2(\nabla_{\mu}\nabla_{\nu}\phi)^3 \right ] + G^{\mu\nu}\nabla_{\mu}\nabla_{\nu}\phi G_5(X,\phi) \, ,
\end{eqnarray}
where $X \equiv -g^{\mu\nu}\nabla_{\mu}\phi\nabla_{\nu}\phi/2$, $(\nabla_{\mu}\nabla_{\nu}\phi) \equiv \nabla_{\mu}\nabla_{\nu}\phi\nabla_{\mu}\nabla^{\nu}\phi$, $(\nabla_{\mu}\nabla_{\nu}\phi)^3 \equiv \nabla_{\mu}\nabla_{\nu}\phi\nabla^{\mu}\nabla^{\rho}\phi\nabla^{\nu}\nabla_{\rho}\phi$, the functions $K(X,\phi)$, $G_3(X,\phi)$, $G_4(X,\phi)$, $G_5(X,\phi)$ are free, and $_,X$ denotes differentiation with respect to $X$. \\

The mimetic version of the above Horndeski model has been studied in a variety of papers recently (e.g. \cite{arroja,rabochaya,bartolo}). We report some particular cases taken into consideration. We remark that the freedom in the free functions $K$, $G_3$, $G_4$, $G_5$, as well as in the function $b(\phi)$ (which, when $\neq 1$, is equivalent to providing a potential for the mimetic field), results in the possibility of reproducing basically any given expansion scenario of the Universe. A specific model studied in \cite{arroja} is one where the functions take the form:
\begin{eqnarray}
K(X,\phi) = c_2X \quad , \quad G_3(X,\phi) = 0 \quad , \quad G_4(X,\phi) = \frac{1}{2} \quad , \quad G_5(X,\phi) = 0 \, .
\end{eqnarray}
In this case, on a flat FLRW background the solution is given by:
\begin{eqnarray}
a(t) = t^{\frac{2}{3(1+w)}} \quad , \quad \phi(t) = \pm\sqrt{-\frac{\alpha}{c_2}} \ln \left ( \frac{t}{t_0} \right ) \quad , \quad b(\phi) = -\frac{1}{\dot{\phi}^2} \, ,
\end{eqnarray}
with $t_0$ and integration constant and $\alpha = -8w/3(1+w)^2$. Thus we see that the scenario under consideration has reproduced the expansion history of an Universe filled with a perfect fluid with EoS $w$ \cite{arroja}. Another case considered in \cite{arroja} is the mimetic cubic Galileon model, where the functions take the following form:
\begin{eqnarray}
K(X,\phi) = c_2X \quad , \quad G_3(X,\phi) = \frac{2c_3}{\tilde{\Lambda}^3X} \quad , \quad G_4(X,\phi) = \frac{1}{2} \quad , \quad G_5(X,\phi) = 0 \, ,
\end{eqnarray}
where the cut-off scale is subsequently set to $\tilde{\Lambda} = 1$. It is then found that the model can reproduce the expansion history of an Universe filled with non-relativistic matter, followed by a cosmological constant dominated expansion analogous to the late-time acceleration we are experiencing \cite{arroja}. The case of a non-minimal coupling to the auxiliary metric $\tilde{g}_{\mu \nu}$ was examined as well, which we will not report on here, and for which we will redirect the reader to the original paper \cite{arroja}. \\

To conclude, we report on the following specific case of mimetic Horndeski model which was studied in \cite{rabochaya}. The Horndeski part of the action of the theory is given by:
\begin{eqnarray}
I_H = \int d^4x \sqrt{-g} \ \left [ \alpha (	XR + (\Box \phi)^2 - \nabla_{\mu}\nabla_{\nu}\phi\nabla^{\mu}\nabla^{\nu}\phi ) + \gamma \phi G_{\mu \nu}\nabla^{\mu}\nabla^{\nu}\phi - \beta\phi\Box\phi \right ] \, ,
\end{eqnarray}
which corresponds to the following choice for the functions discussed previously:
\begin{eqnarray}
K(X,\phi) = 0 \quad , \quad G_3(X,\phi) = \beta\phi \quad , \quad G_4(X,\phi) = \alpha X \quad , \quad G_5(X,\phi) = 0 \, ,
\end{eqnarray}
A number of solutions including cosmological bounces, inflation unified with late-time acceleration, and future singularities, have been discussed. As a specific example, on a flat FLRW background, the following choice of potential for the mimetic field \cite{rabochaya}:
\begin{eqnarray}
V(\phi) = -\beta + 2V_0 + \frac{3V_0^2}{c_1}(\phi - \phi_0)^2 \, ,
\end{eqnarray}
gives rise to the following solution:
\begin{eqnarray}
H(t) = \frac{V_0}{c_1}(t - t_0) \, ,
\end{eqnarray}
which represents a regular bounce solution. Another bounce solution can be obtained by considering the following potential \cite{rabochaya}:
\begin{eqnarray}
V(\phi) = b^2c_1\frac{\sinh^2 b\phi}{\cosh^2 b\phi} \, ,
\end{eqnarray}
for which the scale factor and correspondingly the Hubble parameter read:
\begin{eqnarray}
a(t) = a_0\cosh bt \quad , \quad H(t) = b\frac{\sinh bt}{\cosh bt} \, .
\end{eqnarray}
In Section \ref{mimeticno} we will present a detailed discussion on a specific mimetic Horndeski model, constructed as an extension of a covariant Ho\v{r}ava-like theory of gravity.

\subsection{Einstein-aether theories, Ho\v{r}ava-Lifshitz gravity, and covariant renormalizable gravity}

In closing, we comment on some connections between mimetic gravity and other theories of modified gravity, such connections having been identified recently: namely, the scalar Einstein-aether theory, Ho\v{r}ava-Lifshitz gravity, and a covariant realization of the latter, i.e. covariant renormalizable gravity.

\subsubsection{Einstein-aether theories}

An interesting connection which can be identified is that between mimetic gravity and Einstein-aether theories \cite{ea1,ea2} (see also \cite{ea3,ea4,ea5,jacobson1,jacobson2,jacobson3,jacobson4,jacobson5,jacobson6}). These are a class of Lorentz-violating generally covariant extensions of GR. By Lorentz-violating generally covariant we mean that Lorentz invariance is preserved at the level of the action, only to be broken dynamically. In fact, the theory contains an unit timelike vector $u^{\mu}$ (which is called the aether) whose norm is fixed by a Lagrange multiplier term in the action. This entails the fixing of a preferred rest frame at each spacetime point. In fact, mimetic gravity itself dynamically violates Lorentz symmetry because the gradient of the mimetic field fixes a preferred direction in spacetime. \\

To be precise, mimetic gravity is in correspondence with a particular version of the Einstein-aether theory, namely the scalar Einstein-aether theory \cite{harko,sepangi}. In this theory, the role of aether is played by the gradient of a scalar field, precisely as occurs in mimetic gravity: the role of aether is played by the four-velocity vector which in turn is the gradient of the mimetic field. It should be noted that the scalar Einstein-aether theory is quite different from the original vector theory. Moreover, in the case where the potential for the potential for the scalar field in such theories (corresponding to the potential for the mimetic field in mimetic gravity) is constant, the model corresponds to the IR limit of projectable Ho\v{r}ava-Lifshitz gravity, which we will comment on further in Section \ref{hl}. \\

\subsubsection{Ho\v{r}ava-Lifshitz gravity}
\label{hl}

Recall that Ho\v{r}ava-Lifshitz gravity \cite{horava} (HLG hereafter) is a framework and candidate theory of quantum gravity, wherein gravity is made power-counting renormalizable by altering the graviton propagator in the UV. This is achieved by abandoning Lorentz symmetry as a fundamental symmetry of nature, in favour of a Lifshitz anisotropic scaling in the UV. For an incomplete list of references concerning further work in Ho\v{r}ava-Lifshitz gravity, see for instance \cite{horava1,horava2,horava3,horava4,horava5,horava6,horava7,horava8,horava9,horava10,horava11,horava12,
horava13,horava14,horava15,horava16,horava17,horava18,horava19,horava20,horava21,horava22,horava23,horava24,horava25} and references therein. If the lapse function in HLG is only a function of time, that is, $N = N(t)$, the theory takes the name of projectable Ho\v{r}ava-Lifshitz gravity. Recall that in the Arnowitt-Deser-Misner decomposition of spacetime \cite{arnowittdesermisner}:
\begin{eqnarray}
ds^2 = -N^2c^2dt^2 + g_{ij}(dx^i + N_idt)(dx^j + N^jdt) \, ,
\end{eqnarray}
the function $N$ takes the name of lapse. \\

Previous work has shown that the IR limit of the nonprojectable version of Ho\v{r}ava-Lifshitz gravity can be obtained from the Einstein-aether theory (the vector version) by requiring that the aether be hypersurface orthogonal, i.e. $u_{\mu} = N\nabla_{\mu}T$, where $T$ is a scalar field and $N$ is chosen in such a way to ensure $u_{\mu}$ has unit norm. In this case $T$ is responsible for the preferred foliation in Ho\v{r}ava gravity, and $N$ is the lapse function. If $T$ is set equal to coordinate time, the resulting action is that of nonprojectable Ho\v{r}ava gravity \cite{horava19,jacobson5}. \\

Further requiring that $NdT = S$, where $S$ is a scalar field, reduces the vector Einstein-aether theory to the scalar Einstein-aether theory \cite{jsc}. This condition implies that $N = N(T)$, which upon identification of $T$ with coordinate time corresponds exactly to the defining condition for projectable Ho\v{r}ava gravity. In this case, the unit norm constraint cannot be solved for a generic $N$, but has to be imposed at the level of the action, for instance via a Lagrange multiplier term $\propto \lambda(\nabla_{\mu}S\nabla_{\mu}S - 1)$. Therefore, the condition under which $dS = NdT$ is invariant is that $S$ be invariant under a shift symmetry: $S \rightarrow S + dS$, which shows why the equivalence between scalar Einstein-aether theory and mimetic gravity fails if a nonzero potential for the mimetic field is included \cite{jsc}. Notice also that, as is known, dark matter emerges as an integration constant in the IR limit of projectable Ho\v{r}ava-Lifshitz gravity \cite{horava16}. Given the correspondence between this theory and mimetic gravity, then, it should not come as a surprise that dark matter emerges in a similar fashion within the framework of mimetic gravity, as a purely geometrical effect. \\

A complete proof of the equivalence between mimetic gravity and the IR limit of projectable Ho\v{r}ava-Lifshitz gravity was presented in \cite{matarrese}. In particular, it was shown that the action for the IR limit of projectable Ho\v{r}ava-Lifshitz gravity can be written as:
\begin{eqnarray}
S = S_{\text{EH}} + \int d^4x \sqrt{-g} \ \left [ \frac{\Sigma}{2}(g^{\mu \nu}\partial_{\mu}\phi\partial_{\nu}\phi + 1) + \frac{\gamma}{2}(\Box \phi)^2 \right ] \, ,
\end{eqnarray}
which we immediately recognize as the action for mimetic gravity without potential, with the addition of a higher derivative term which, as we shall see shortly, was actually added shortly after the original formulation of the theory on purely phenomenological grounds, to cure some undesirable properties at the perturbation level. In Ho\v{r}ava-Lifshitz gravity, $\Sigma$ is a Lagrange multiplier which had been introduced to enforce the projectability condition. This equivalence was demonstrated rigorously in Appendix A of \cite{matarrese}. Therefore, mimetic gravity appears in the IR limit of a candidate theory of quantum gravity. \\

\subsubsection{Covariant renormalizable gravity}

Recall that HLG achieves power-counting renormalizability by breaking diffeomorphism invariance. However, this breaking appears explicitly at the level of the action. This has been at the center of criticism, which have related this explicit breaking to the appearance of unphysical modes in the theory which couple strongly in the IR \cite{horava12,horava13,horava19,horava20,horava21}. Therefore, in order to circumvent these issues it would be desirable to have a theory which preserves the renormalizability properties of Ho\v{r}ava gravity in the UV, but retains diffeomorphism invariance at the level of the action. Therefore, diffeomorphism invariance should be broken dynamically in the UV. \\

An example of such theory has been presented by Nojiri and Odintsov \cite{horava18,horava23,horava24} (see also \cite{crg1,crg2,crg3,crg4,crg5,crg6,crg7,crg8,crg9}), and comes under the name of covariant renormalizable gravity (CRG henceforth). The theory features a non-standard coupling of a perfect fluid to gravity. When considering perturbations around a flat background, this non-standard coupling dynamically breaks diffeomorphism invariance. The price to pay is the presence of this exotic fluid, which could have a stringy origin. In particular, one formulation of CRG introduces the fluid via a Lagrange multiplier term, precisely as done in mimetic gravity. For this reason, in the limit where the non-standard coupling of the fluid to gravity disappears, one recovers the action of mimetic gravity. In one of its equivalent formulations, the action of CRG is given by \cite{horava18,horava23,horava24}:
\begin{eqnarray}
I &=& \int d^4x \ \sqrt{-g} \left \{ \frac{R}{2} - \alpha \left [ \left ( \partial ^{\mu}\phi\partial ^{\nu}\phi \nabla _{\mu}\nabla _{\nu} + 2U_0\nabla ^{\rho}\nabla _{\rho} \right ) ^n \left ( \partial ^{\mu}\phi\partial ^{\nu}\phi R_{\mu \nu} + U_0R \right ) \right ] ^2 \right.\nonumber\\&&\left.- \lambda \left ( \frac{1}{2}\partial _{\mu}\phi \partial ^{\mu}\phi + U_0 \right ) \right \} \, ,
\end{eqnarray}
where $U_0$ and $\alpha$ are constants, and $U_0 = 1/2$ to recover the mimetic constraint. In particular, the above case corresponds to the $z = 2n+2$ version of CRG, where $z$ is the dynamical critical exponent which quantifies the degree of anisotropy between space and time in the UV regime of the theory. \\

Following the initial proposal by Nojiri and Odintsov, other CRG-like models were studied in recent years. For instance, one particular CRG-like model was studied by Cognola et al. in \cite{crg4}, where black hole and de Sitter solutions were also studied. The action of such CRG-like theory takes the following form \cite{crg4}:
\begin{eqnarray}
I = \frac{1}{2}\int d ^4x\sqrt{-g} \left \{ R - 2\Lambda - \alpha \left [ \left ( R ^{\mu \nu} - \frac{R}{2}g ^{\mu \nu} \right ) \nabla _{\mu}\phi\nabla _{\nu}\phi \right ] ^n - \frac{\lambda}{2} \left ( g ^{\mu \nu}\nabla _{\mu}\phi\nabla _{\nu}\phi + 1 \right )  - V (\phi) \right \} \ .
\label{actionpotential}
\end{eqnarray}
In the above action, the Horndeski-like coupling of the scalar field to curvature (for the case $n = 1$) has been considered copiously in the literature, for an incomplete list see for instance \cite{horndeski1,horndeski7,horndeski14,
coupling1,coupling2,coupling3,coupling4,coupling6,coupling7,coupling8,coupling9,coupling10} and references therein. \\

In the continuation of our review, after a brief interlude on perturbations in mimetic gravity, we shall consider a case study of a mimetic-like theory. Our choice of case study will fall upon the CRG-like model of Cognola et al. as defined by the action in Eq.(\ref{actionpotential}). We shall study in detail its cosmological solutions, and perturbations around a flat background. A pathological behaviour of the model when considering perturbations around a flat FLRW background will be cured by appropriately modifying the model.

\section{A brief interlude: cosmological perturbations in mimetic gravity}

Before proceeding to the case study of a specific mimetic-like model, it is mandatory to discuss the issue of perturbations in mimetic gravity. Recall in Section \ref{MG}, we remarked that mimetic gravity does not properly belong to the class of modified theories of gravity with a full extra scalar degree of freedom. Instead, the additional scalar is constrained by the Lagrange multiplier, a situation quite different from, e.g., that of Horndeski models which have a proper scalar degree of freedom. \\

However, it is clear that the Lagrange multiplier kills the wave-like parts of the scalar degree of freedom: in other words, given that the constraint takes out any higher derivative, it is not possible to have oscillating (wave-like solutions). As a consequence, we can already envisage that the sound speed in the minimal mimetic gravity model will satisfy $c_s = 0$. This implies, as we have anticipated, that there are no propagating scalar degrees of freedom in the theory. \\

The property of vanishing sound speed in mimetic gravity can be rigorously demonstrated, by considering small longitudinal perturbations around a flat background \cite{m2}. The line element is then given by:
\begin{eqnarray}
ds^2 = -( 1+2\Phi(t, {\bf x}) )dt^2 + a^2(t)( 1-2\Psi(t, {\bf x}) )\delta_{ij}dx^i dx^j \, , \quad i,j = 1,2,3 \, ,
\label{NG}
\end{eqnarray}
$\Phi\equiv\Phi(t, {\bf x})$ and $\Psi\equiv\Psi(t, {\bf x})$ being functions of the space-time coordinates such that $|\Phi(t,x)|$, $|\Psi(t,x)|$ $\ll 1$, and $g^{00}(t,x)\simeq-1+2\Phi(t,x)$, $g^{11}(t,x)\simeq a(t)^{-2}(1+2\Psi(t,x))$. 
Here, we used the conformal Newtonian gauge. Moreover, from field equations we also have:
\begin{eqnarray}
\Phi(t, {\bf x})=\Psi(t, {\bf x})\,.
\label{109}
\end{eqnarray}
Correspondingly, given that around a flat FLRW background the mimetic field plays the role of ``clock'', we perturb it as:
\begin{eqnarray}
\phi = t + \delta \phi(t, {\bf x}) \,,
\label{pertphi}
\end{eqnarray}
where $|\delta\phi|\equiv|\delta \phi(t,x)|\ll 1$,
which together with Eq.(\ref{norm}) implies that:
\begin{eqnarray}
\Phi = \delta \dot{\phi} \,.
\label{phiPhi}
\end{eqnarray}
Then, by perturbing the $0i$ components of the Einstein equations, a relatively straightforward calculation \cite{m2} shows that the evolution equation for the perturbation to the mimetic field, $\delta \phi$, satisfies the following equation:
\begin{eqnarray}
\ddot{\delta \phi} + H\dot{\delta \phi} + \dot{H}\delta \phi = 0 \, .
\label{p}
\end{eqnarray}
The aforementioned pathological property of perturbations in mimetic gravity can be noticed from Eq.(\ref{p}). The evolution equation for perturbations to the mimetic field does not depend on the Laplacian (or in general on spatial derivatives) of the latter. In other words, there is no term of the form $c_s^2\Delta \delta \phi$ in Eq.(\ref{p}), which implies that the sound speed is identically zero. This means that, even when pressure is nonvanishing, the dust degree of freedom induced in mimetic gravity behaves as dust with zero sound speed, and as such quantum perturbations to the mimetic field cannot be defined in the usual fashion. Else, they would fail in providing the seeds for the observed large-scale structure of the Universe which grow via gravitational instability. We remark once more that this behaviour is not unexpected, given that the condition enforced by the Lagrange multiplier eliminates wave degrees of freedom. Moreover, this fact has been shown in all generality for mimetic Horndeski models in \cite{bartolo}. \\

In order to have a theory whose quantum perturbations can be defined in a sensible way, the minimal action for mimetic gravity has to be modified, for instance by introducing higher derivative (HD) terms. As an example, consider the following action \cite{m2}:
\begin{eqnarray}
I = d^4x \sqrt{-g} \ \left [ R + \lambda(g^{\mu \nu}\partial_{\mu}\phi\partial_{\nu}\phi + 1) - V(\phi) + \frac{1}{2}\gamma (\Box \phi)^2 \right ] \, .
\end{eqnarray}
The corresponding equations of motion read:
\begin{eqnarray}
G^{\mu}_{\nu} = \left [ V + \gamma \left ( \partial_{\alpha}\phi\partial^{\alpha}\chi \right ) \right ]\delta^{\mu}_{\nu} + 2\lambda \partial^{\mu}\phi\partial_{\nu}\phi - \gamma(\partial^{\mu}\chi\partial_{\nu}\phi + \partial^{\mu}\phi\partial_{\nu}\chi ) \, ,
\end{eqnarray}
where $\chi \equiv \Box \phi$. Thus, identification of the mimetic field with time on a flat FLRW background implies that $\chi = 3H$. There are two main effects of the introduction of the specific HD term on the theory: one at the level of the background and one at the level of perturbations. At the level of the background, it can easily be shown that the Friedmann equation is modified from Eq.(\ref{friedmann}) to \cite{m2}:
\begin{eqnarray}
2\dot{H} + 3H^2 = \frac{2}{2-3\gamma}V \, ,
\end{eqnarray}
which corresponds to a renormalization of the amplitude of the potential. At the level of perturbations, it can be shown that Eq.(\ref{p}) is modified to the following \cite{m2}:
\begin{eqnarray}
\ddot{\delta \phi} + H\dot{\delta \phi} - \frac{c_s^2}{a^2}\Delta \delta \phi + \dot{H}\delta \phi = 0 \, ,
\end{eqnarray}
where:
\begin{eqnarray}
c_s^2 = \frac{\gamma}{2 - 3\gamma} \, .
\end{eqnarray}
Therefore, the addition of the higher derivative term results in a small but nonvanishing sound speed, which implies that the behaviour of mimetic matter deviates from the usual perfect fluid dust. The nonvanishing sound speed also results in the possibility of defining quantum perturbations in a sensible way. It is beyond the scope of our review to provide further technical details on the matter, but it can be shown that the simple model we have just described is capable of producing red-tilted (i.e. $n_s < 1$) scalar perturbations which are enhanced over gravity waves (implying a small value of $r$), which is consistent with observations from Planck and BICEP2/Keck Array \cite{m2}. \\

In concluding this brief interlude, let us also spare a few words on further modifications of mimetic gravity involving higher derivative terms. These have been studied in \cite{capela} and \cite{mirzagholi}, in particular by adding the previously discussed $(\Box \phi)^2$ term, as well as a term proportional to $\nabla_{\mu}\nabla_{\nu}\phi\nabla^{\mu}\nabla^{\nu}\phi$. It was noticed that these terms affect the growth of perturbations below the sound horizon, in particular suppressing the growth of those with large momenta. The result is the presence of a cutoff in the matter power spectrum for perturbations below a certain wavelength. On larger scales, instead, the predictions for the matter power spectrum match those of collisionless cold dark matter. \\

The suppression of power on small scales is particularly intriguing in the light of the observation that the collisionless cold dark matter paradigm appears to suffer from a number of shortcomings on subgalactic scales. The core-cusp problem refers to the discrepancy between N-body simulations of collisionless cold dark matter, which predict a cuspy profile for the dark matter halo in galaxies, and observations which instead suggest a cored profile towards the center. The discrepancy is particularly large for dwarf galaxies, but indications that a cored profile is favoured for larger galaxies as well persist (see e.g. \cite{corecusp1,corecusp2,corecusp3,corecusp4,corecusp5}). Moreover, these same simulations predict an abundance of substructure which is approximately 10 times larger than what we actually observe, an issue which is referred to as the ``missing satellites problem'' (see e.g. \cite{missingsatellite1,missingsatellite2,missingsatellite3}). Although this problem is most acute for satellite galaxies, it exists for field galaxies as well (e.g. \cite{zwaan}). To make matters worse, the most massive subhalos in these simulations find no observed counterpart, despite one would expect star formation to be more efficient within them: this is known as the ``too big to fail problem'' (see e.g. \cite{tbtf1,tbtf2,tbtf3,tbtf4}). For an incomplete list of comprehensive reviews on these issues, refer for instance to \cite{peter,popolo} and references therein. \\

Several approaches to solving these problems exist in the literature. If one insists that dark matter is cold and collisionless, then an important role must be played by the baryonic content of the Universe. In fact, it has been argued in several works that baryonic feedback processes (see e.g. \cite{feedback1,feedback2,feedback3,feedback4,feedback5,feedback6,feedback7,feedback8,feedback9,feedback10}) due to supernovae or to the stellar and gaseous content of galaxies, or dynamical friction between dark matter and baryon clumps (see e.g. \cite{friction1,friction2}), can in principle solve or at least alleviate these problems. Alternatively, the small-scale discrepancies might be taken as an indication that something is lacking in the collisionless cold dark matter picture, with DM possibly having sizeable self-interactions. This approach has been undertaken in a number of works, see for instance \cite{dm1,dm2,dm3,dm4,dm5,dm6,dm7,dm8,dm9,dm10} and references therein. In particular, a possibility which has received a lot of attention recently is that where the paucity of structure on small scales is explained by modifying the properties of dark matter in such a way that the resulting matter power spectrum is suppressed at large wavenumbers, by coupling the dark matter to a bath of dark radiation (e.g. a massless or light dark photon) or to a light scalar, which delays kinetic decoupling, refer for instance to \cite{photon1,photon2,photon3,photon4,photon5,photon6,photon7,photon8,photon9,photon10,
photon11,photon12,photon13,photon14,photon15} and references therein. \\

A different mechanism, but with similar outcomes, occurs in mimetic gravity. Namely, the suppression of small-scale power, operated by the higher derivative terms, has the potential to solve the missing satellite problem and the too big to fail problem, as shown in \cite{capela}. Moreover, the same higher derivative terms could potentially also cure the caustic singularities from which mimetic gravity suffers. The reason for this is that the higher derivative terms effectively correspond to terms parametrizing dissipation, or viscosity, which emerges due to the fact that the velocity dispersion of the dust is now non-zero. Actually, in \cite{capela} it was also argued that these same dissipative terms could alleviate the core-cusp problem. Although these discussions remain preliminary, it is very interesting to note that modifying the action for mimetic gravity by the addition of higher derivative terms could provide a solution to the small-scale structure puzzles of collisionless cold dark matter.

\section{A case study: mimetic Horndeski covariant Ho\v{r}ava-like gravity}
\label{mimeticno}

Having discussed in detail the physics behind mimetic gravity, and many of its extensions, we now provide a detailed case study of a specific mimetic model. Our choice falls on the covariant Ho\v{r}ava-like theory of gravity first discussed by Cognola et al. \cite{crg4}, with action defined by Eq.(\ref{actionpotential}), which we will argue can be viewed as a mimetic Horndeski model. We will study the model, its background solutions, and scalar perturbations in detail. Not unexpectedly, we will find that the sound speed of the minimal model is vanishing, and thus quantum perturbations cannot be defined in the usual way. To circumvent this problem, we modify the model by the addition of higher order terms, repeating the analysis of background solutions and scalar perturbations, and show that it will be necessary to go beyond the Horndeski framework in order to have a nonvanishing sound speed. The discussion in this section is largely based on the work of \cite{cognola}.

\subsection{Mimetic covariant Ho\v{r}ava-like gravity}

Let us start from the action of the CRG-like model first discussed by Cognola et al. \cite{crg4}, which is given by the following:
\begin{eqnarray}
I= 
\frac{1}{2}\,\int d^4x\,\sqrt{-g}\, \left\{R-2\Lambda
-\alpha\left[\left(R^{\mu\nu}-\frac\beta2\,Rg^{\mu\nu}\right)\nabla_\mu\phi\nabla_\nu\phi\right]^n
    -\lambda\,\left(\frac12\,g^{\mu\nu}\partial_\mu\phi\partial_\nu\phi+U_0\right)\right\}\ ,
\label{lcrg}
\end{eqnarray}
where $\alpha,\beta$ are arbitrary constants, $\Lambda$ is the cosmological constant, $n$ is a natural number, $\lambda$ is a Lagrange multiplier, and $U_0$ determines the constraint imposed on the gradient of the cosmological field $\phi$. Recall that the Horndeski-like coupling in the action above, for $n=1$, has been considered in several works, e.g. \cite{horndeski1,horndeski14,
coupling1,coupling2,coupling3,coupling4,coupling5,coupling6,coupling7,coupling8,coupling9,coupling10}. If we set $U_0=1/2$, we see that the constraint on the gradient of the scalar field corresponds precisely to that of mimetic gravity, Eq.(\ref{norm}). Thus we can extend the model to include the mimetic field by adding a potential for the scalar field, which is now made dynamical. The action now reads:
\begin{eqnarray}
I = \frac{1}{2}\int d ^4x\sqrt{-g} \left \{ R - 2\Lambda - \alpha \left [ \left ( R ^{\mu \nu} - \frac{R}{2}g ^{\mu \nu} \right ) \nabla _{\mu}\phi\nabla _{\nu}\phi \right ] ^n - \frac{\lambda}{2} \left ( g ^{\mu \nu}\partial _{\mu}\phi\partial _{\nu}\phi + 1 \right )  - V (\phi) \right \} \,,
\label{actionpotential1}
\end{eqnarray}
where we have set the dimensionless parameter $\beta = 1$, and added a potential for the mimetic field, $V(\phi)$. Variation with respect to $\lambda$ immediately leads to Eq.(\ref{norm}), whereas variation respect to the field modifies Eq.(\ref{phieq}) to:
\begin{eqnarray}
\frac{ d V(\phi)}{d\phi} & = & \nabla _{\mu} \left [ \left ( 2n\alpha F ^{n-1} G ^{\mu \nu} + \lambda g ^{\mu \nu} \right ) \partial _{\nu}\phi  \right ] \nonumber\\
& = & \frac{1}{\sqrt{-g}}\partial _{\mu} \left \{ \sqrt{-g} \left [ \left ( 2n\alpha F ^{n-1}G ^{\mu \nu} + \lambda g ^{\mu \nu} \right ) \partial _{\nu}\phi \right ] \right \} \,,
\label{equationscalar}
\end{eqnarray}
where we define the following quantities:
\begin{eqnarray}
F \equiv T_{\mu \nu}R^{\mu \nu}-\frac{RT}{2} \, , \qquad T_{\mu \nu} \equiv \nabla_{\mu} \phi \nabla_{\nu} \phi \, , \qquad T \equiv g^{\mu \nu}T_{\mu \nu}=-1 \, .\label{FT}
\end{eqnarray}
Finally, variation of the action with respect to the metric leads to the gravitational field equations, which for this theory read:
\begin{eqnarray}
G_{\mu \nu}+\Lambda g_{\mu \nu} + \frac{\alpha}{2}F^ng_{\mu \nu} & = & n\alpha F^{n-1} \left [ R^{\rho}_{\mu}T_{\rho \nu} + R^{\rho}_{\nu}T_{\rho \mu} - \frac{1}{2} \left ( TR_{\mu \nu} + RT_{\mu \nu} \right ) \right ] + \frac{\lambda}{2}T_{\mu \nu} \nonumber \\
&&\hspace{-2cm} +n\alpha \left [ D_{\alpha \beta \mu \nu}(T^{\alpha \beta}F^{n-1}) - \frac{1}{2}D_{\mu \nu} \left ( TF^{n-1} \right ) \right ] + \Omega ^{\alpha \beta}\frac{\delta T_{\alpha \beta}}{\delta g^{\mu \nu}} - g_{\mu \nu} \frac{V(\phi)}{2} \, ,
\label{gmunu}
\end{eqnarray}
where we have defined the differential operators:
\begin{eqnarray}
D_{\alpha \beta \mu \nu} & \equiv & \frac{1}{4} \left [ \left ( g_{\mu \alpha}g_{\nu \beta} + g_{\nu \alpha}g_{\mu \beta} \right ) \Box + g_{\mu \nu} \left ( \nabla_{\alpha}\nabla_{\beta}+\nabla_{\beta}\nabla_{\alpha} \right ) 
\right.
\nonumber\\
&&\left.
- \left ( g_{\mu \alpha}\nabla_{\beta}\nabla_{\nu} + g_{\nu \alpha}\nabla_{\beta}\nabla_{\mu} + g_{\mu \beta} \nabla_{\alpha}\nabla_{\nu} + g_{\nu \beta}\nabla_{\alpha}\nabla_{\mu} \right ) \right ] \, , \nonumber \\
D_{\mu \nu} & \equiv & g_{\mu \nu}\Box - \frac{1}{2} \left ( \nabla_{\mu}\nabla_{\nu}+\nabla_{\nu}\nabla_{\mu} \right ) \,,
\end{eqnarray}
$\Box\equiv \nabla^i\nabla_i$ being the d'Alambertian operator. Note that in the above Eq.(\ref{gmunu}), $\Omega_{\mu \nu}$ is a tensor that will not play any role if $T_{\mu \nu}$ does not depend on the metric, which we assume is our case, thus we can drop it from the gravitational field equations. Finally, the form of the Lagrange multiplier $\lambda$ can be determined from the trace of Eq.(\ref{gmunu}):
\begin{eqnarray}
- R + 4\Lambda - \frac{\lambda}{2}T &=&\nonumber\\
&&\hspace{-3cm} 2\alpha F ^n(n-1) + \frac{n\alpha}{2} \left ( g _{\mu \nu}\Box + \nabla _{\mu}\nabla _{\nu} + \nabla _{\nu}\nabla _{\mu} \right ) \left ( T ^{\mu \nu}F ^{n-1} \right ) - \frac{3n\alpha}{2}\Box \left ( TF ^{n-1} \right ) - 2V(\phi) \, .
 \label{trace}
\end{eqnarray}
We will consider the case with $n=1$, which is particularly interesting given that it is equivalent to a specific instance of a mimetic Horndeski model.

\subsubsection{Cosmological solutions}

Let us now consider cosmological solutions, in particular consideing a flat FLRW metric, Eq.(\ref{FRWmetric}). If we take the hypersurfaces of constant time to be equal to those of constant $\phi$, by making use of the mimetic constraint (\ref{norm}), we see that the field can be identified (up to an integration constant) with time:
\begin{eqnarray}
\phi = t \,.
\end{eqnarray}
As in the original mimetic gravity, the scalar field can induce an effective cold dark matter component given that, from Eq.(\ref{phieq}), one has $-(G-T)\propto a^3$, and recall that $-(G-T)$ gives the energy density of the perfect dust-like fluid induced by mimetic field (or, more precisely, by its gradient which plays the role of 4-velocity field). However, the introduction of additional terms depending on the field in the action Eq.(\ref{actionpotential1}) change these features. Here, we would like to study the behaviour of mimetic field and the cosmological solutions of the more involved theory (\ref{actionpotential1}). \\

Let us begin by considering the tensor $T_{\mu \nu}$ in Eq.(\ref{FT}), which reads:
\begin{eqnarray}
T _{00} = \dot{\phi} ^2 = 1 \, , \quad T_{0i} = T_{i0} = T_{ij}=0\, , \quad i,j=1,2,3\,.
\end{eqnarray}
The, from Eq.(\ref{equationscalar}) we derive:

\begin{eqnarray}
\frac{1}{a ^3}\partial _0 \left[ a ^3 \left ( 2n\alpha (3H ^2) ^n - \lambda  \right ) \right] = \frac{d V(\phi)}{d\phi} \,.
\label{consn}
\end{eqnarray}
Given the mimetic constraint Eq.(\ref{norm}), this equation is a consequence of the two EOMs inferred from Eq.~(\ref{gmunu}):
\begin{eqnarray}
0 & = & \Lambda - 3H^2 + \frac{\alpha}{2}(1-4n)(3 H^2)^n + \frac{\lambda}{2} + \frac{V(\phi)}{2} \, , \\
0 & = & \Lambda - 3H^2 - 2\dot H + \frac{\alpha}{2}(1-2n)(3 H^2)^n + 3^{n-1}\alpha n(1-2n)\dot H H^{2n-2} + \frac{V(\phi)}{2} \,.
\label{m}
\end{eqnarray}
In our analysis, as the second independent equation, together, with Eq.(\ref{consn}), we choose the trace equation Eq.(\ref{trace}), namely:
\begin{eqnarray}
\frac{\lambda}{2} = 6\dot{H} + 12H ^2 - 4\Lambda + \alpha (5n-2)(3H ^2) ^n + 3 ^nn\alpha (2n-1)H ^{2n-2}\dot{H} - 2V(\phi)\,.
\label{tracef}
\end{eqnarray}
For the simplest case $V(\phi)=0$, from Eq.(\ref{consn}) we get:
\begin{eqnarray}
\left [ 2n\alpha(3H^2)^n - \lambda \right ] = \frac{C_0}{a^3} \, ,
\label{friedmannno}
\end{eqnarray}
where $C_0$ is an integration constant. Thus, in the limit $\alpha=0$, we recover the original mimetic gravity model of \cite{m1}, while when $\alpha\neq 0$ we may interpret this equation as a generalized Friedmann equation, with $C_0$ determining the amount of dark matter in the universe (given that its contribution scales with scale factor $a$ as $a^{-3}$, as is expected for dust). Moreover, in the limits given by:
\begin{eqnarray}
1 \ll \alpha\,, \quad n = 1\,,  \nonumber\\
\frac{1}{\alpha^{n-1}}\ll R \ , \quad 1 < n\,,
\end{eqnarray}
if we neglect the Cosmological Constant $\Lambda$, as well as the integration constant $C_0$, one gets from Eq.(\ref{tracef}):
\begin{eqnarray}
-\frac{\dot H}{H^2}=\frac{2}{n}\,.
\end{eqnarray}
Thus, if $1/\alpha^{n-1}$ is set at the scale of the early-time acceleration, models with $1\ll n$ lead to cosmological solutions for inflation. \\

In the case where we set $n=1$, the parameter $\alpha$ is dimensionless and the EOMs are second order, which is not surprising given that our model is a special case of Horndeski's theory. Therefore, we can incorporate the Cosmological Constant in the potential and obtain from Eqs.(\ref{consn}, \ref{tracef}):
\begin{eqnarray}
\frac{dV}{dt} & = &\frac{1}{a ^3}\partial _0 \left[ a ^3 \left (6\alpha H ^2 - \lambda   \right ) \right] \,,
\label{l}\\
\lambda & = & 6(2+\alpha)\dot{H} + 6 ( 4+3\alpha )H^2-4V(\phi) \,,
\label{t1}
\end{eqnarray}
where we have taken into account that $V\equiv V(\phi)=V(t)$ on the FLRW space-time, so that we can replace the potential derivative of the field with its time derivative. \\

Let us manipulate Eq.(\ref{m}) further, which gives:
\begin{eqnarray}
 2\dot{H}+3 H^2=\frac{V}{(2+\alpha)} \,,
\label{z}
\end{eqnarray}
showing that the model we are considering is essentially equivalent to the model proposed in \cite{m2}. Since Eq.~(\ref{l}) is a consequence of Eqs.(\ref{t1}\,,\ref{z}), we may choose to infer the cosmological solutions from (\ref{z}) only. This equation is a non-linear Riccati type equation and can be transformed in the linear second order differential equation :
\begin{eqnarray}
\ddot{u} - \frac{3}{4(2+\alpha)}V u=0 \,,
\label{z1}
\end{eqnarray}
by introducing the Sturm-Liouville canonical substitution
\begin{eqnarray}
H = \frac{2}{3}\frac{\dot{u}}{u} \,,
\quad a = u^{2/3} \,.
\end{eqnarray}
Let us discuss some examples by starting from Eq.(\ref{z1}). If $V=V_0$ is constant, we recover the de Sitter solution:
\begin{eqnarray}
u\sim \exp[3H_0 t/2]\,,\quad a\sim \exp[H_0 t]\,,\quad H_0=\frac{2}{3}\sqrt{\frac{3V_0}{4(2+\alpha)}}\,.
\end{eqnarray}
Another well-motivated choice for the potential is a quadratic one:
\begin{eqnarray}
V(\phi)=3(2+\alpha)\left [ H_0^2+\beta ^2 \left ( 2\phi-\phi_0\right)\left(-H_0+\frac{\beta^2}{4}(2\phi-\phi_0) \right ) - \frac{2}{3}\beta ^2  \right] \, ,
\end{eqnarray}
where $H_0$ is a constant Hubble parameter and $\beta,\phi_0$ are dimensional constants (in the specific case $[\beta]=[\phi_0^{-1}]=[H]$). After the identification $\phi=t$ the explicit solution for $u$ is found to be:
\begin{eqnarray}
u(t)=u_0 e^{\frac{3}{2}H_0 t-\frac{3\beta^2}{4} t(t-2 t_0)}\,,\quad t_0\equiv\frac{\phi_0}{2} \, ,\label{39}
\end{eqnarray}
with $u_0$ constant and $t_0$ a fixed time. The Hubble parameter is given by:
\begin{eqnarray}
H \equiv \frac{2}{3}\frac{\dot{u}}{u} = H_0 - \beta ^2 \left ( t-t_0\right ) \,,
\label{starobinsky}
\end{eqnarray}
and we see that for $t$ close to $t_0$, one has a quasi-de Sitter expansion, while for large $t_0 \ll t$, the Hubble parameter tends to vanish. This solution corresponds to a Starobinsky-like accelerated expansion \cite{inf2} in the Jordan frame and gives an interesting inflationary solution. \\

Let us provide one final example with a potential given by:
\begin{eqnarray}
V(\phi) = \frac{4 A^2(2+\alpha)}{3}\frac{\cosh A \phi}{1+\cosh A \phi} \, ,
\end{eqnarray}
where $0<A$ is a constant with mass-dimension 1. The corresponding solution is given by:
\begin{eqnarray}
u(t) = 1+\cosh A t \,, \quad a=(1+\cosh A t )^{\frac{2}{3}}\,,\quad
H=\frac{2A}{3}\frac{\sinh At}{(1+\cosh A t )}\,.\label{bouncesol}
\end{eqnarray}
This solution represents a cosmological bounce with $-\infty<t\,,\phi<+\infty$ and shows that the mimetic field may act as a phantom fluid.   

\subsubsection{Cosmological scalar perturbations}

In this section, we will consider the scalar perturbations around the FLRW metric Eq.(\ref{FRWmetric}) in the model defined by Eq.(\ref{actionpotential1}) with $n=1$, analyzed in the previous section. The perturbed metric in conformal Newtonian gauge is given by Eq.(\ref{NG}). Concerning perturbations to the mimetic field, we obtain once more Eq.(\ref{pertphi}) which leads to Eq.(\ref{phiPhi}) once more. Note that in this case the identity Eq.(\ref{109}), which is valid for the original mimetic dark matter model, is no longer true. Furthermore, we notice that:
\begin{eqnarray}
T_{00} = 1 + 2\delta \dot{\phi} \, , \quad T_{0i} = \partial_ i \delta\phi \, , \quad T = -1 + {\cal O}(\Phi^2) \, .\label{Tpert}
\end{eqnarray}
From the $(1,2)$-component of Eq.(\ref{gmunu}), we obtain:
\begin{eqnarray}
G_{12} \left ( 1 - \frac{\alpha}{2} \right ) = \alpha D_{\alpha \beta 12}T^{\alpha \beta} \,,
\end{eqnarray}
with:
\begin{eqnarray}
G_{12} = -\partial _x\partial _y ( \Phi - \Psi ) \, , \quad D_{\alpha \beta 12}T^{\alpha \beta} = H\partial _x\partial _y \delta \phi + \partial _x\partial _y \delta \dot{\phi} \,.
\end{eqnarray}
Therefore, we obtain:
\begin{eqnarray}
\Psi=\Phi+\left(\frac{2\alpha}{2-\alpha}\right)(H\delta\phi+\delta\dot\phi)\,.
\end{eqnarray}
From (0,1)-component of Eq.(\ref{gmunu}) we derive:
\begin{eqnarray}
G_{01} \left ( 1 + \frac{\alpha}{2} \right ) = \alpha \dot{H} \partial _x\delta\phi + \frac{\lambda}{2} \partial _x\delta\phi + \alpha D_{\alpha \beta 01}T^{\alpha \beta} \, ,
\end{eqnarray}
with:
\begin{eqnarray}
G_{01}=2\partial _x \left ( \dot\Psi + H\Phi \right ) \, , \quad D_{\alpha \beta 01}T^{\alpha \beta} = -(H^2 + \dot{H}) \partial _x\delta\phi \, , \quad \lambda = 6\alpha H^2 - 4\dot{H} - 2\alpha\dot{H} \,,
\end{eqnarray}
where the last equality is a consequence of Eqs.(\ref{t1},\ref{z}). \\

Finally, we can obtain a closed equation for $\delta\phi$, which reads:
\begin{eqnarray}
\delta \ddot{\phi} + H \delta \dot{\phi} + \dot{H} \delta\phi = 0 \,,\label{52}
\end{eqnarray}
From the above we immediately read that the sound speed is vanishing, given that there is no dependence on the Laplacian of $\delta \phi$. The implications are that in these kinds of models, scalar perturbations do not propagate (as in the original mimetic model of \cite{m2}), rendering the usual definition of quantum perturbations quite problematic. Given that this feature is rooted in the model being of the mimetic Horndeski form \cite{bartolo}, to address the problem, either we must take $n\neq 1$ in Eq.(\ref{actionpotential}) or we must modify the original action along the lines of \cite{m2,capela}.

\subsection{Modified higher order mimetic Horndeski model}

With our goal being that of addressing the problem of scalar perturbations, we modify the model given by Eq.(\ref{actionpotential1}) for $n=1$ as:
\begin{eqnarray}
I = \frac{1}{2}\int d ^4x\sqrt{-g} \left [ R(1 + a g^{\mu \nu}\nabla_{\mu}\phi\nabla_{\nu}\phi) - \frac{c}{2}(\Box \phi)^2 + \frac{b}{2} (\nabla_{\mu}\nabla_{\nu}\phi)^2 - \frac{\lambda}{2} \left ( g ^{\mu \nu}\nabla _{\mu}\phi\nabla _{\nu}\phi + 1 \right )  - V (\phi) \right ] \,.
\label{actionpotential2}
\end{eqnarray}
The original model given by Eq.(\ref{actionpotential}) for $n=1$ is recovered for $a=\alpha/2$ and $b=c=2\alpha$ by using the following identity \cite{kobayashi}:
\begin{eqnarray}
-\frac{1}{2}g^{\mu \nu}\nabla_{\mu}\phi\nabla_{\nu}\phi R + (\Box \phi)^2 - (\nabla_{\mu} \nabla_{\nu}\phi)^2 = G^{\mu \nu}\nabla_{\mu}\phi\nabla_{\nu}\phi + \mbox{total derivative} \, .
\end{eqnarray}
On the other hand, for generic values of $a, b, c$, the action Eq.(\ref{actionpotential2}) describes  a higher order derivative model in the scalar sector with field equations at the fourth order, namely:
\begin{eqnarray}
(1-a)G_{\mu \nu} & = & \frac{1}{2}g_{\mu \nu} \left [ \frac{b}{2}\phi^{\alpha \beta}\phi_{\alpha \beta} - \frac{c}{2}(\Box\phi)^2 -V(\phi) \right ] + \lambda \nabla_{\mu}\phi\nabla_{\nu}\phi \nonumber \\
& - & b \phi_{\mu \rho}\phi^{\rho}_{\nu} + \frac{b}{2}g^{\alpha \beta} \left [ \nabla_{\alpha}(\phi_{\mu \nu} \nabla_{\beta}\phi) - \nabla_{\alpha}(\phi_{\mu \beta}\nabla_{\nu}\phi) - \nabla_{\alpha}(\phi_{\nu \beta}\nabla_{\mu}\phi) \right ] \nonumber \\
& + & c \left [ \phi_{\mu \nu} + g_{\mu \nu}g^{\alpha \beta}\nabla_{\alpha}(\Box\phi\nabla_{\beta}\phi) - \nabla_{\mu} \Box\phi \nabla_{\nu}\phi - \nabla_{\nu}\Box\phi \nabla_{\mu}\phi \right ] \, ,
\label{fieldmod}
\end{eqnarray}
where we adopted the notation:
\begin{eqnarray}
\phi_{\alpha \beta} \equiv \nabla_{\alpha}\nabla_{\beta}\phi\,. 
\end{eqnarray}
Let us explore some cosmological applications. On a flat FRW metric, Eq.(\ref{FRWmetric}), Eq.(\ref{z}) is modified as follows:
\begin{eqnarray}
2\dot{H} + c_h H^2 = c_v V(\phi) \,,
\label{bulk}
\end{eqnarray}
where:
\begin{eqnarray}
c_h \equiv \frac{12a + 3b - 9c - 12}{4a + b - 3c - 4} \, , \quad c_v \equiv \frac{2}{4 - 4a - b + 3c} \, . 
\end{eqnarray}
It is understood that by setting $a =\alpha/2$ and $b = c = 2 \alpha$, we recover Eq.(\ref{z}). Similarly to how we proceeded in the previous Section, we can rewrite Eq.(\ref{bulk}) as:
\begin{eqnarray}
\ddot{u} - \frac{c_h c_v V(\phi)}{4} u=0\,,
\label{z2}
\end{eqnarray}
where we introduced the auxiliary function $u$:
\begin{eqnarray}
H = \frac{2}{c_h}\frac{\dot{u}}{u} \,, \quad a(t) = u^{\frac{2}{c_h}} \,,
\end{eqnarray}
By choosing the following potential:
\begin{eqnarray}
V(\phi)=\frac{9}{c_h c_v}\left [ H_0^2+\beta ^2 \left ( 2\phi-\phi_0\right)\left(-H_0+\frac{\beta^2}{4}(2\phi-\phi_0) \right ) - \frac{2}{3}\beta ^2  \right] \, ,
\end{eqnarray}
with $H_0\,,\beta \, ,\phi_0$ dimensional constants, we recover the Starobinsky-like solution Eqs.(\ref{39},\ref{starobinsky}). If we choose the following potential:
\begin{eqnarray}
V(\phi) = \frac{4 A^2 }{c_vc_h}\frac{\cosh A \phi}{1+\cosh A \phi} \,,
\end{eqnarray}
$0<A$ being a constant, we recover the bounce solution Eq.(\ref{bouncesol}). Therefore, we can recover all the solutions of the Horndeski-like model previously analyzed. We will now study the scalar perturbations around FRW space-time.

\subsubsection{Cosmological scalar perturbations}

If we consider the perturbed metric, Eq.(\ref{NG}), and the perturbed field, Eq.(\ref{pertphi}), we still recover Eqs.(\ref{phiPhi},\ref{Tpert}). Therefore, from the $(i, j)$-components, $i,j=1,2,3$ of Eq.(\ref{fieldmod}) we obtain now:
\begin{eqnarray}
\Psi=\Phi+\frac{b}{2-2 a}\left(\delta \dot{\phi}+H\delta\phi \right)\,.
\label{r2}
\end{eqnarray}
Moreover, from the components $(0, i)$ or $(i, 0)$, $i=1,2,3$ of Eq.(\ref{fieldmod}) we derive:
\begin{eqnarray}
\delta \ddot{\phi} + H \delta \dot{\phi}-\frac{c_s^2}{a^2}\nabla^2 \delta \phi + \dot{H}\delta\phi=0\,,
\end{eqnarray}
where the non-vanishing squared sound speed $c^2_s$ reads
\begin{eqnarray}
c^2_s \equiv \frac{b-c}{2 c_2} \,,\quad c_2 \equiv \frac{(2+b-2a)(4+3c-4a-b)}{4(a-1)} \,.
\end{eqnarray}
We immediately note that $c_s^2=0$ when $b=c$ and we recover Eq.(\ref{52}). Since only for $b=c$ does the model fall within the Horndeski class, we can consider $(b-c)$ as being the Horndeski breaking parameter. This also confirms that, in order to obtain a non-vanishing sound speed, it is necessary to go beyond the mimetic Horndeski framework, i.e. $b \neq c$. The result is analogous to that obtained in \cite{m2, capela}.

\section{Spherically symmetric solutions in mimetic gravity}

In this section, we will explore static spherically symmetric solutions (SSS) in mimetic gravity. To do so, let us return to the general formulation of mimetic gravity with action given by Eq.(\ref{actionlm}). Variation of the action respect to the metric with the mimetic constraint Eq.(\ref{norm}) simply leads to:
\begin{eqnarray}
G_{\mu \nu}& = &  \frac{\lambda}{2} g^{\mu\nu}\partial_\mu\phi\partial_\nu\phi- g_{\mu \nu} \frac{V(\phi)}{2} \,,\label{EOM1gen}
\end{eqnarray}
namely Eq.(\ref{gmunu}) with $\alpha=0$ and $\Lambda=0$. The trace of this equation reads:
\begin{eqnarray}
R = \frac{\lambda}{2} + 2V(\phi)\,. 
\label{lambda}
\end{eqnarray}
Thus, from (\ref{EOM1gen}), we get:
\begin{eqnarray}
G _{\mu \nu} = (R - 2V(\phi))\partial _{\mu}\phi\partial _{\nu}\phi - \frac{1}{2}g _{\mu \nu}V(\phi) \ .
\label{EOM1}
\end{eqnarray}
The continuity equation of the mimetic field is derived as:
\begin{eqnarray}
\frac{1}{\sqrt{-g}}\partial _{\nu} \left ( \sqrt{-g}\lambda\partial ^{\nu}\phi \right ) =\frac{ dV(\phi)}{d\phi}\,.
\label{fieldeq} 
\end{eqnarray}
which is automatically satisfied when (\ref{EOM1}) holds true.

In this chapter we will consider pseudo-SSS space-times, whose general topological formulation is given by:
\begin{eqnarray}
ds^2=-a(r)^2 b(r) dt^2+\frac{dr^2}{b(r)}+r^2\left(\frac{d\rho^2}{1-k\rho^2}+\rho^2 d\phi^2\right) \,,
\label{metric}
\end{eqnarray}
where $a(r),b(r)$ are functions of the radial coordinate $r$ and the manifold is a sphere when $k=1$, a torus when $k=0$ or a compact hyperbolic manifold when $k=-1$. The Ricci scalar in this case reads:
\begin{eqnarray}
R = -\frac{1}{r^2}\left[3r^2\,b'(r)\frac{a'(r)}{a(r)}+r^2 b''(r)+2r^2\,b\left(r\right)\frac{a''(r)}{a(r)}+4r\,b'(r)+4 r b(r)\,\frac{a'(r)}{a(r)}+2b(r)-2k\right] \,,
\label{SSSRicci}
\end{eqnarray}
where the prime denotes a derivative with respect to $r$. The symmetries of the EOMs require the mimetic field to be a function of $r$ only, namely $\phi\equiv\phi(r)$, and from Eq.(\ref{norm}) one has:
\begin{eqnarray}
\phi'(r)=\sqrt{-\frac{1}{b(r)}} \,,
\label{fieldSSS}
\end{eqnarray}
leading to a pure imaginary expression for the field, which is to be expected from a time-like vector $\partial_\mu\phi$ with temporal component equal to zero. Therefore, it is clear that the correspondence with the dark matter is only formal, which justifies the introduction of the potential to make the mimetic field dynamical and possibly reproduce the dark matter phenomenology, since the four-velocity vector $u_\mu$ in (\ref{stresspf}) cannot be physical. This has been explored in \cite{sebastianicurves} and will be discussed in Section \ref{curves}. \\

The (0,0)- and (1,1)-components of the field equations (\ref{fieldeq}) lead to:\footnote{
To derive the EOMs, one may plug the expression for the Ricci  scalar directly into the action to obtain, after integration by parts \cite{mioLagr1}:
\begin{eqnarray*}
\mathcal L=2a\left(1-r b'(r)-b(r)\right)+a(r)r^2\lambda\left(b(r)\phi'(r)^2+1\right)-a(r)r^2V(\phi)\,.
\end{eqnarray*}
Thus, the derivatives respect to $a(r)$ and $b(r)$ with (\ref{norm}) lead to (\ref{uno}) and (\ref{duebis}), while the derivation respect to $\phi$ leads to (\ref{fieldeqSSS}).
}
\begin{eqnarray}
k-b'(r)r-b(r)=\frac{V(\phi)r^2}{2} \ ,
\label{uno}
\end{eqnarray}
\begin{eqnarray}
\left(b'(r)r+2r \frac{a'(r)}{a(r)} b(r)+b(r)-k\right)=\frac{\lambda}{2} b(r) r^2\phi'(r)^2-\frac{V(\phi)r^2}{2} \,,
\label{due}
\end{eqnarray}
where $\lambda$ is given by Eq.(\ref{lambda}). \\

We can also rewrite Eq.(\ref{due}) with (\ref{fieldSSS})--(\ref{uno}) as:
\begin{eqnarray}
4a'(r) b(r)=-\lambda a(r) r \ .
\label{duebis}
\end{eqnarray}
Finally, from Eq.(\ref{fieldeq}), one finds the following:
\begin{eqnarray}
\frac{d}{d r}\left(a(r)b(r)\lambda r^2\phi'\right)=a(r)r^2\frac{d V(\phi)}{d\phi} \ .
\label{fieldeqSSS}
\end{eqnarray}
As a first investigation, we will consider the case where potential is equal to zero, that is, vacuum solutions.

\subsection{Vacuum solutions}

In this subsection we set $V(\phi)=0$. From Eq.(\ref{uno}) we immediately see that:
\begin{eqnarray}
b(r)=\left ( k - \frac{r_s}{r}\right) \,,
\label{bex1}
\end{eqnarray}
$r_s$ being a mass scale, either positive or negative, while the second field equation (\ref{due}) with (\ref{lambda}, \ref{SSSRicci}) leads to:
\begin{eqnarray}
a(r)=a_1+\frac{a_2}{\sqrt{1-\frac{k r_s}{r}}}\left[\left(\sqrt{1-\frac{k r_s}{r}}\right)\log\left[\sqrt{\frac{r}{r_0}}\left(1+\sqrt{1-\frac{k r_s}{r}}\right)\right]-1\right] \,,\quad k=\pm 1\,,
\label{aex1}
\end{eqnarray}
\begin{eqnarray}
a(r)=a_1+a_2\left[2\left(\frac{r}{r_0}\right)^{3/2}+3\right]\,,
\end{eqnarray}
$a_1, a_2$ being constants and $r_0$ a length scale. If $a_2=0$, namely $R=0$ and $\lambda=0$,  we recover the topological Schwarzschild solution of General Relativity.  When $a_2\neq 0$, we can pose $a_1=0$. Thus, the solution for flat topology ($k=0$) is as follows:
\begin{eqnarray}
ds^2=-a_2^2\left[2\left(\frac{r}{r_0}\right)^{3/2}+3\right]^2\left(\frac{\tilde r_s}{r}\right)dt^2+
\frac{dr^2}{\left(\frac{\tilde r_s}{r}\right)}+r^2\left(d\rho^2+\rho^2 d\phi^2\right)\,,\quad 0<\tilde r_s\,,
\end{eqnarray}
where $\tilde r_s=-r_s$ has to be positive to preserve the metric signature. The Ricci scalar is non-zero and reads:
\begin{eqnarray}
R=-\frac{6\tilde r}{2r^3+3\left(\frac{r}{r_0}\right)^{3/2}r_0^3}\,.
\end{eqnarray}
This solution presents a naked singularity at $r=0$, as for the corresponding topological case $k=0$ of the Schwarzschild metric. \\

The spherical case ($k=1$) is the more interesting and is given by:
\begin{eqnarray}
ds^2=-a_2^2\left[
\left(\sqrt{1-\frac{r_s}{r}}\right)
\log\left[\sqrt{\frac{r}{r_0}}\left(1+\sqrt{1-\frac{r_s}{r}}\right)\right]-1\right]^2 dt^2
+\frac{dr^2}{\left(1-\frac{r_s}{r}\right)}+r^2\left(d\theta^2+\sin^2\theta d\phi^2\right)\,,
\label{k1}
\end{eqnarray}
where we have introduced the polar coordinates $\theta\,,\phi$ in the angular part (see also Ref.~\cite{deruelle}). In this case the Ricci scalar reads:
\begin{eqnarray}
R=-\frac{1}{r^2\left[
\left(\sqrt{1-\frac{r_s}{r}}\right)
\log\left[\sqrt{\frac{r}{r_0}}\left(1+\sqrt{1-\frac{r_s}{r}}\right)\right]-1\right]}\,.
\end{eqnarray}
For $r_s<0$ the metric is regular everywhere, but when $0<r_s$, the solution is regular and preserves the signature only in the region $r_s<r$ the solution is regular and preserves the signature, since for $r<r_s$ the metric coefficient $g_{00}(r)\equiv-a^2(r) b(r)$ acquires an imaginary part. In this respect, we note that the special choice 
$r_s=r_0$ when $r<r_s$ leads to:
\begin{eqnarray}
g_{00}(r)\equiv a(r)^2 b(r)=-a_2^2\left(\sqrt{\frac{r_s}{r}-1}\arctan\left[\sqrt{\frac{r_s}{r}-1}\right]+1\right)^2\,,
\end{eqnarray}
which is real and negative. \\

In general, we observe that when $0<r_s$, the point $r=r_s$ cannot represent an horizon like for the Schwarzschild space-time, due to the fact that $g_{11}(r)\equiv 1/b(r)$ diverges but $g_{00}(r)=-a_2^2$ and the Ricci scalar are regular (we also cannot associate to $r=r_s$ any thermodynamical quantity such as temperature $T=(a(r)/2) d b(r)/dr$ which diverges at $r=r_s$); moreover, for $r<r_s$ the metric becomes imaginary (or, in the special case $r_s=r_0$, acquires the signature (--++)) and does not allow a wormhole description. \\

Finally, the topological case $k=-1$ reads:
\begin{eqnarray}
ds^2=-a_2^2 \left[
\left(\sqrt{1+\frac{r_s}{r}}\right)
\log\left[\sqrt{\frac{r}{r_0}}\left(1+\sqrt{1+\frac{r_s}{r}}\right)\right]-1\right]^2 dt^2
+\frac{dr^2}{\left(-1-\frac{r_s}{r}\right)}+r^2\,\left(\frac{d\rho^2}{1+\rho^2}+\rho^2 d\phi^2\right)\,,\label{k-1}
\end{eqnarray}
and the Ricci scalar is given by:
\begin{eqnarray}
R=-\frac{1}{r^2\left[
\left(\sqrt{1+\frac{r_s}{r}}\right)
\log\left[\sqrt{\frac{r}{r_0}}\left(1+\sqrt{1+\frac{r_s}{r}}\right)\right]-1\right]}\,.
\end{eqnarray}
In this case, for $0<r_s$, the metric coefficient $g_{11}(r)\equiv 1/b(r)$ is negative and the solution is unphysical. On the other hand, for $r_s<0$, we obtain $0<g_{11}(r)$ when $r<-r_s$, but $g_{00}(r)\equiv -a^2(r) b(r)$ becomes imaginary (except for the special choice $r_0=-r_s$).

\subsection{Non-vacuum solutions}

In this subsection, we will consider the case $V(\phi)\neq 0$. We immediately see that for $V(\phi)=2\Lambda$ with $\Lambda$ a cosmological constant, one solution of Eqs.(\ref{uno},\ref{due}) with $\Lambda=0$ is the topological 
Schwarzschild de Sitter solution:
\begin{eqnarray}
b(r)=k-\frac{r_s}{r}-\frac{\Lambda r^2}{3}\,,\quad a(r)=a_1\,,\label{SdSsol}
\end{eqnarray}
$r_s\,,a_1$ being constants. \\

In considering other solutions, we will take the spherical case $k=1$ in Eq.(\ref{metric}). By using Eq.(\ref{fieldSSS}) one readily obtains from Eqs.(\ref{duebis},\ref{fieldeqSSS}):
\begin{eqnarray}
\phi(r)=\pm i\int\frac{dr}{\sqrt{b(r)}}\,,\quad 4\frac{d}{dr}\left(a'(r) b(r)^{3/2}r\right)=a(r) r^2\sqrt{b(r)}\frac{d V(r)}{d r} \,,
\label{equations}
\end{eqnarray}
where we treat the potential as a function of $r$. These two equations with Eq.(\ref{uno}) can be used to reconstruct the potential when a choice for $b(r)$ is made. We will now provide some examples of the reconstruction technique. \\ 

Let us consider a linear modification to the Schwarzschild metric:
\begin{eqnarray}
b(r)=\left ( 1 - \frac{r_s}{r} + \gamma r \right ) \ ,
\label{bex2}
\end{eqnarray}
$r_s,\gamma$ being constants whose mass-dimension is positive, such that from  Eq.(\ref{uno}) we obtain,:
\begin{eqnarray}
V(r)=-\frac{4\gamma}{r} \,.
\end{eqnarray}
The corresponding solution for the mimetic field is found to be an elliptic function, and explicit expressions can be given only in limiting cases. When $r \approx r_s$, one can neglect the linear correction in Eq.(\ref{bex2}) to recover the solution in Eqs.(\ref{bex1},\ref{aex1}) with $k=1$. Thus, the mimetic field reads:
\begin{eqnarray}
\phi(r\ll\sqrt{r_s/\gamma})\simeq\pm i\left[r\sqrt{1-\frac{r_s}{r}}+\frac{r_s}{2}\log\left[2r\left(1+\sqrt{1-\frac{r_s}{r}}\right)-r_s\right]\right] \,,
\label{24}
\end{eqnarray}
or, by expanding the result around $r=r_s$:
\begin{eqnarray}
\phi(r\simeq r_s)\simeq \phi_s\pm2i\sqrt{r_s(r-r_s)}\,, \quad r\simeq r_s-\frac{(\phi_s-\phi)^2}{4r_s} \ ,
\label{25}
\end{eqnarray}
with $\phi_s=\pm (i r_s/2)\log(r_s)$. The explicit form of the potential at $r\simeq r_s$, $\phi\simeq\phi_s$ is found to be:
\begin{eqnarray}
V(\phi\simeq\phi_s)\simeq-\frac{4\gamma}{r_s}-\frac{\gamma(\phi_s-\phi)^2}{r_s^3} \ .
\label{Vex2}
\end{eqnarray}
On the other hand, for large distances, we may ignore the Newtonian term in Eq.(\ref{bex2}) and from the second equation in Eq.(\ref{equations}) we derive:
\begin{eqnarray}
a(\sqrt{r_s/\gamma}\ll r)\simeq\frac{c_1(4+6\gamma r)+3c_2\sqrt{1+\gamma r}-c_2(2+3\gamma r)\arctan[\sqrt{1+\gamma r}]}
{\sqrt{1+\gamma r}} \ ,,
\end{eqnarray}
with $c_{1,2}$ dimensional constants. We can choose $c_1=1/4$ and $c_2=0$, and in the given limit the metric simply reads:
\begin{eqnarray}
ds^2(\sqrt{r_s/\gamma}\ll r)\simeq-\left(1+\frac{3\gamma r}{2}\right)^2dt^2+\frac{dr^2}{(1+\gamma r)}+r ^2\left(d\theta^2+\sin^2\theta d\phi^2\right) \,.
\label{m22}
\end{eqnarray}
The corresponding expressions for the field and the potential are given by:
\begin{eqnarray}
\phi(r)\simeq\pm\frac{2i\sqrt{1+r\gamma}}{\gamma}\,,\quad r\simeq-\frac{4+\gamma^2\phi^2}{4\gamma}\,,\quad V(\phi)\simeq\frac{16\gamma^2}{4+\gamma^2\phi(r)^2} \ .
\label{Vex2bis}
\end{eqnarray}
Here, we must require $4/\gamma^2<|\phi|^2$ in order to guarantee the positivity of $r$. The behaviour of the potential can be derived by interpolating the expressions in Eqs.(\ref{Vex2}, \ref{Vex2bis}). \\

The metric under investigation reduces to the usual Schwarzschild space-time for short distances, while at large distances its 00-component behaves as:
\begin{eqnarray}
g_{00}(\sqrt{r_s/\gamma}\ll r) \simeq-\left(1+\frac{3\gamma r}{2}\right)^2\,.
\end{eqnarray}
Therefore, the corresponding Newtonian potential $\Phi(r)=-(g_{00}(r)+1)/2$ acquires linear and quadratic contributions with respect to the Newtonian solution. The quadratic correction can be viewed as a negative cosmological constant in the background and can be ignored if $\gamma^2 r^2$ is sufficiently small. On the other hand, the linear term (for $\gamma > 0$) could help in explaining the inferred flatness of galactic rotation curves, which has been interpreted as one of the key evidences for the presence of dark matter. We will return to this issue in the next Section and analyse the problem more closely there, where we will provide a metric whose cosmological constant-like contribution is independent from the linear one. \\ 

As a second example of reconstruction procedure, we consider the following ansatz:
\begin{eqnarray}
b(r)=1-\frac{r_s^2}{r^2} \ ,
\end{eqnarray}
$r_s$ being once more a positive dimensional constant. From Eq.(\ref{uno}) we get:
\begin{eqnarray}
V(r)=-\frac{2r_s^2}{r^4} \,,
\end{eqnarray}
and the metric is fixed by making use of the second equation in Eqs.(\ref{equations}) which leads to:
\begin{eqnarray}
a(r)=\frac{\left[c_1+c_2\arctan\left[\frac{r}{\sqrt{r_s^2-r^2}}\right]\right]}{\sqrt{1-\frac{r_s^2}{r^2}}} \,.
\end{eqnarray}
The metric signature for $r_s<r$ is preserved when $c_2=0$, and by choosing $c_1=1$ we obtain
\begin{eqnarray}
ds^2=-dt^2+\frac{dr^2}{\left(1-\frac{r_s^2}{r^2}\right)}+r ^2\left(d\theta^2+\sin^2\theta d\phi^2\right) \,.
\label{mex3}
\end{eqnarray}
The field and the potential are found to be:
\begin{eqnarray}
\phi=\pm i r\sqrt{1-\frac{r_s^2}{r^2}}\,,\quad r=\sqrt{r_s^2-\phi^2}\,,\quad V(\phi)=-\frac{2r_s^2}{(r_s^2-\phi^2)^2} \,.
\end{eqnarray}
The radial coordinate is always real and positive.

The above metric [Eq.(\ref{mex3})] is very interesting, given that it may be used to describe a traversable wormhole (see e.g. \cite{wormholes1,wormholes2,wormholes3,wormholes4,wormholes5,wormholes6,wormholes7,wormholes8,wormholes9,wormholes10,
wormholes11,wormholes12,wormholes13,wormholes14,wormholes15,wormholes16,wormholes17,wormholes18,wormholes19,wormholes20,
wormholes21,wormholes22,wormholes23,wormholes24,wormholes25} and references therein, see also \cite{modern1,modern2,modern3,modern4,modern5,modern6,modern7,modern8,modern9,modern10,
modern11,modern12,modern13,modern14,modern15} for recent work on the subject), where the space is divided in two spherical holes connected by a ``throat'' located at $r=r_s$. Moreover, the metric satisfies the following requirements:
\begin{enumerate}
\item $g_{00}(r)$ and $g_{11}^{-1}(r)$ are well defined for all $r_s\leq r$;
\item $g_{00}(r)$ is regular on the throat with 
$g_{00+}(r_s)=g_{00-}(r_s)$ and
$g_{00+}'(r_s)=g_{00-}'(r_s)$;
\item $g_{11}^{-1}(r_s)=0$ and $0<g_{11}^{-1}(r)$ for all $r_s<r$;
\item Given $g_{11}(r)^{-1}=[1-\tilde b(r)/r]$, we have $\tilde b_+'(r_s)=\tilde b'_{-}(r_s)<1$.
\end{enumerate}
The above correspond to the traversability conditions \cite{wormholes3,wormholes6}, and thus we conclude that the wormhole described by our solution is traversable. We also note that our space-time is asymptotically flat. \\

In the next Section we will use a different approach to fix $a(t)^2 b(r)$, which encodes the physical Newtonian potential, in order to apply mimetic gravity to reproduce the phenomenology of galactic rotation curves.

\section{Rotation curves of galaxies in mimetic gravity}
\label{curves}

So far we have seen that, within mimetic gravity, dark matter emerges as a geometrical effect at a cosmological level, i.e. in the form of a perfect fluid whose energy density decays as $a^{-3}$. However, one of the first (and probably most renown) clues as to the presence of dark matter came not from cosmological scales, but from astrophysical ones. In fact, it was the observations of Vera Rubin, along with her colleagues Kent Ford, David Burstein, and Norbert Thonnard, that galactic rotation curves were asimptotically flat or even slightly growing with radius $r$, far beyond the region where luminous matter is present, that signalled the presence of something unexpected \cite{du4,du5,du6}. In fact, if luminous matter were the only responsible for the shape of the inferred rotation curves, simple classical mechanics calculations dictate that such curve should fall as $v_{\text{rot}}(r) \propto 1/\sqrt{r}$, which clearly did not match what was observed. Therefore, it would be interesting to see whether it is possible, within mimetic gravity, to explain the shape of galactic rotation curves, thus fulfilling the requirements of a successful theory of dark matter on both cosmological and astrophysical scales. \\

The first solution to this problem was found in \cite{sebastianicurves}. The key idea is closely related to that of MOND \cite{mond}: to introduce a new scale in the theory, relating it to either the scale where predictions of Newtonian gravity fail, or to a scale intrinsically present in the rotation curves data. The work of \cite{sebastianicurves} has drawn from two classes of examples in the literature: the static solution of conformal Weyl gravity \cite{riegert} used in such context in \cite{mannheim,mannheimrev,mannheimlast,dwarf}, and solutions in $f(R) = R^n$ gravity \cite{capo}, used in \cite{salucci,salucci2,salucci3}. Let us begin by considering the first cases, which features linear and quadratic corrections to the Schwarszchild metric.

For our purpose, it is convenient to redefine:
\begin{eqnarray}
\tilde a(r)=a(r)^2 b(r)\,,
\end{eqnarray}
within the metric (\ref{metric}) with $k=1$, namely:
\begin{eqnarray}
ds^2=-\tilde a(t)dt^2 + \frac{dr^2}{b(r)}+r^2\left(r^2 d\theta^2+\sin^2\theta d\phi^2\right) \ .
\label{metricrc}
\end{eqnarray}
Since:
\begin{eqnarray}
a(r)=\sqrt{\frac{\tilde a(r)}{b(r)}}\,,\quad a'(r)=\frac{1}{2\sqrt{\tilde a(r) b(r)}}\left(\tilde a'(r)-\tilde a (r)\frac{b'(r)}{b(r)}\right)\,,
\end{eqnarray}
by using Eq.(\ref{uno}) the second in Eqs.(\ref{equations}) reads:
\begin{eqnarray}
\frac{d}{dr}\left[\left(\tilde a'(r)b(r)-\tilde a(r)b'(r)\right)\frac{r}{\sqrt{\tilde a(r)}}\right]=
\sqrt{\tilde a(r)}\left[-b''(r)r-\frac{2}{r}\left(1-b(r)\right)\right]\,.\label{eqb}
\end{eqnarray}
We choose the following ansatz for $\tilde a(r)$:
\begin{eqnarray}
\tilde a(r)=1-\frac{r_s}{r}+\gamma r-\lambda_0 r^2 \ ,
\label{aex}
\end{eqnarray}
with $r_s,\lambda_0,\gamma$ positive dimensional constants. Given the metric element $g_{00}(r)=-\tilde a(r)$, the Newtonian potential reads:
\begin{eqnarray}
\Phi(r)=-\frac{(g_{00}(r)+1)}{2}\,.
\label{Np0}
\end{eqnarray}
Thus, the Newtonian potential associated to Eq.(\ref{aex}) can be described in the following way:
\begin{enumerate}
\item at small distances, the metric leads to a classical Newtonian term $r_s/r$;
\item at very large distances, the ``cosmological constant'' term $\lambda_0 r^2$ emerges, reflecting the fact that the metric is immersed in a cosmological de Sitter background;
\item at intermediate distance the linear term $\gamma r$ appears, revealing a new feature respect to the Schwarzschild-de Sitter metric at galactic scales.
\end{enumerate}
At intermediate galactic scales, we can safely assume that the $\lambda_0 r^2$ term is negligible, hence the Newtonian potential Eq.(\ref{Np0}) reads:
\begin{eqnarray}
\Phi(r)\simeq-\frac{r_s}{2r}\left(1-\frac{\gamma r^2}{r_s}\right) \ .
\label{Np}
\end{eqnarray}
Correspondingly, the rotational velocity profile reads:
\begin{eqnarray}
v_{\text{rot}}^2 \simeq v_{\text{Newt}} ^2 + \frac{\gamma c^2r}{2} \, ,
\end{eqnarray}
where we reintroduced the speed of light $c$ and $v_{\text{Newt}}$ is the contribution expected from the luminous matter component. Therefore, on sufficiently large scales, $v_{\text{rot}}$ does not fall-off as per the Keplerian result $1/\sqrt{r}$, but increases slightly as $\sqrt{r}$. This is particularly true for galaxies where the falling Newtonian contribution cannot compete with the rising determined by the $\gamma$ term (depending of course on the size of $\gamma_0$), which occurs for small and medium sized low surface brightness (LSB) galaxies. In fact rotation curve results for such galaxies exhibit precisely this behaviour (see e.g. discussion in \cite{mannheimlast} in the context of conformal Weyl gravity): that is, the rotation curves of these galaxies start rising immediately. \\

The situation is different for sufficiently extended galaxies, for instance large high surface brightness (HSB) galaxies. For these galaxies the Newtonian contribution in might be sufficient to complete with the rising linear term, $\propto \gamma r$. This leads to a region of approximate flatness before any rise starts, and is consistent with the data for such galaxies (see e.g. \cite{mannheimlast}). Moreover, for these galaxies $r$ might be sufficiently large that the de Sitter term $\propto r^2$ should be taken into account. Because of the negative sign, the effect of this term is to reduce the velocity, thus the rotational velocity profile is given by:
\begin{eqnarray}
v ^2 \simeq v _{\text{Newt}} ^2 + \frac{\gamma c ^2r}{2} - \lambda _0c ^2r ^2 \ .
\label{vquad}
\end{eqnarray}
Clearly, sufficiently far from the center of such galaxies, the quadratic term takes over and arrests the rising behaviour driven by the linear term. This is in perfect agreement with data from HSB galaxies, which are large enough to feel the effect of the de Sitter term. Moreover, the negative sign in front of this term has another important implication: given that $v^2$ cannot go negative, bound orbits are no longer possible on scales greater than $R \sim \gamma _0/2\lambda _0$. This could provide a dynamical explanation for the maximum size of galaxies, determined by the interplay between the linear ($\gamma$) and the quadratic ($\lambda _0$) terms. \\

Let us turn to the question of reproducing such behaviour in mimetic gravity. In order to reconstruct the complete form of the metric (\ref{metricrc}), we must use
Eq.(\ref{eqb}) to derive:
\begin{eqnarray}
b(r)=\frac{\left(1-\frac{r_s}{r}+\gamma r-\lambda_0 r^2\right)\left(1-\frac{3r_s}{r}+\frac{\gamma r}{3}+\frac{c_0}{r^2}\right)}{\left(1-\frac{3 r_s}{2 r}+\frac{\gamma r}{2}\right)^2} \ ,
\label{bex}
\end{eqnarray}
with $c_0$ a constant. The (on-shell) form of the potential inferred from Eq.~(\ref{uno}) is quite involved
and results to be:
\begin{eqnarray}
V(r)&=&-\frac{2}{3r^2(2r-3r_s+\gamma r^2)^3}[54 r_s^2 r-27r_s^3+
171\gamma r_s^2 r^2-8\gamma^2\lambda_0 r^7+r^4(16\gamma+7r_s\gamma^2+324r_s\lambda_0)
\nonumber\\&& 
+4 r_s r^3(-17\gamma-108r_s\lambda_0)+
r^6(\gamma^3-44\gamma\lambda_0)+6r^5(\gamma^2-12\lambda_0+12r_s\gamma\lambda_0)
\nonumber\\&&
-12c_0[-r_s+2r(1+r_s\gamma)+2r^3(\gamma^2+\lambda_0)-\gamma\lambda_0 r^4+3r^2(\gamma-3r_s\lambda_0)]] \,.
\label{Vrot}
\end{eqnarray}
Moreover, one uses the first equation in (\ref{equations}) to recover the behaviour of the field which can be found only in the limiting cases \cite{sebastianicurves}. \\

In the limit $\gamma=\lambda_0=0$ (corresponding to small distances), one has:
\begin{eqnarray}
\tilde a(r)\simeq1-\frac{r_s}{r}\,,\quad b(r)\simeq\frac{4(c_0+r(r-3r_s))(r-r_s)}{r(2r-3r_s)^2} \ .
\label{45}
\end{eqnarray}
Thus, if we set:
\begin{eqnarray}
c_0=\frac{9r_s^2}{4}\,,
\label{c0}
\end{eqnarray}
we have:
\begin{eqnarray}
\tilde a(r)\simeq 1-\frac{r_s}{r}\,,\quad b(r)\simeq 1-\frac{r_s}{r} \,,
\end{eqnarray}
namely we recover the vacuum Schwarzschild solution of General Relativity. Analogously to Eqs.(\ref{24},\ref{25}) we derive:
\begin{eqnarray}
\phi(r\simeq r_s)\simeq \phi_s\pm2i\sqrt{r_s(r-r_s)}\,,\quad
r\simeq r_s-\frac{(\phi_s-\phi)^2}{4r_s} \,,
\end{eqnarray}
with $\phi_s=\pm (i r_s/2)\log[r_s]$. In this case the potential behaves as given in the following:
\begin{eqnarray}
V(\phi\simeq\phi_s)\simeq-\frac{32\gamma}{3r_s}+\frac{13\gamma(\phi-\phi_s)^2}{r_s^3} \ .
\end{eqnarray}
In the limit $r_s=\gamma=c_0=0$ (cosmological scales) we obtain:
\begin{eqnarray}
\tilde a(r)=b(r)\simeq(1-\lambda_0 r^2) \,,
\label{50}
\end{eqnarray}
which corresponds to the static patch of the de Sitter solution. Then, the field assumes the form:
\begin{eqnarray}
\phi\simeq\pm i\frac{\text{arcsin}\left[\sqrt{\lambda_0} r\right]}{\sqrt{\lambda_0}}\,,\quad
r\simeq\pm\frac{\sin\left[\sqrt{\lambda_0}|\phi|\right]}{\sqrt{\lambda_0}}\,.
\end{eqnarray}
We note that $0<r$ as long as $0<r<H_0^{-1}$, where $H_0^{-1}=1/\sqrt{\lambda_0}$ is the cosmological horizon of the de Sitter solution with positive cosmological constant. The corresponding behaviour of the potential is:
\begin{eqnarray}
V(\phi)\simeq 6\lambda_0\mp\frac{4\gamma}{3}\left(\frac{\sqrt{\lambda_0}}{\sin\left[\sqrt{\lambda_0}|\phi|\right]}+ 4\sqrt{\lambda_0}\sin\left[\sqrt{\lambda_0}|\phi|\right]\right)\,.
\end{eqnarray}
This result with $\gamma=0$ is consistent with \cite{m2}. \\

Finally, in the limit $r_s=\lambda_0=c_0=0$ of galactic scales, the metric reads:
\begin{eqnarray}
\tilde a(r)\simeq\left(1+\gamma r\right)\,,\quad
b(r)\simeq\frac{4\left(1+\gamma r\right)\left(3+\gamma r\right)}{3\left(2+\gamma r\right)^2}\,.
\end{eqnarray}
In this case, the field is given by:
\begin{eqnarray}
\phi\simeq\pm\frac{i}{2\gamma} \sqrt{3(3+4\gamma r+\gamma^2 r^2)}\,,\quad
r\simeq\frac{-6\mp\sqrt{9-12\gamma^2\phi^2}}{3\gamma}\,,
\end{eqnarray}
while the potential in Eq.(\ref{Vrot}) behaves as:
\begin{eqnarray}
V(r)\simeq-\frac{2\gamma(16+6\gamma r+\gamma^2 r^2)}{3r(2+\gamma r)^3}\,.
\end{eqnarray}
The explicit reconstruction of the potential at the galactic scale must take into account that only the solutions with the positive sign inside $r$ yields a positive radius for $0<\gamma $. Thus, we obtain:
\begin{eqnarray}
V(\phi) = -\frac{2\sqrt{3}\gamma^2\left(27-4\gamma^2\phi^2+2\sqrt{9-12\gamma^2\phi^2}\right)}{\left(3-4\gamma^2\phi^2\right)^{3/2}\left(-6+\sqrt{9-12\gamma^2\phi^2}\right)} \ .
\end{eqnarray}
In conclusion, we are able to describe, in limiting cases the behaviour of the potential $V(\phi)$ which leads to the solution in Eqs.(\ref{aex},\ref{bex}) with the corresponding Newtonian potential in Eq.(\ref{Np}). The considered solution turns out to be the Schwarzschild solution at small distances, the static patch of the de Sitter space-time at cosmological distance and, most intriguingly, presents a linear term at the galactic scales which can address the problem of galactic rotation curves in mimetic gravity. We note that, from the dependence of the potential on the field, the potential is always real, as it is required for consistency. \\

To fix the values of $\gamma,\lambda$ in the metric, one needs the data from rotation curves of galaxies.  Since at the galactic scale our metric reproduces the Newtonian potential (\ref{Np}), while asymptotically turns out to coincide with the de Sitter metric, we can follow the analyses of \cite{mannheimlast,dwarf}, where the same potential has been found as a result of the Riegert solution \cite{riegert} of conformal Weyl gravity. Thus, we adopt the results in \cite{mannheimlast,dwarf}, where the same parameters were fitted to rotation curves. The total sample fitted consists of 138 galaxies, 25 of them being dwarf galaxies. Some of the galaxies in the sample are sufficiently extended as to be sensitive to the de Sitter quadratic term. We refer the reader \cite{mannheimlast,dwarf} and references therein for data on the galaxies included. \\

As per the analysis of \cite{mannheimlast,dwarf} The potential we reconstructed yields an excellent fit to the rotation curves, with a reduced $\chi^2$ of $\approx 1$. In fact, the linear and quadratic corrections to the Newtonian potential capture the falling and rising features in the rotation curves quantitatively rather than barely qualitatively. The best-fit values to our $\gamma$ and $\lambda_0$ parameters in mimetic gravity [Eq.(\ref{aex})] to be \cite{mannheimlast,dwarf}:
\begin{eqnarray}
\gamma \simeq 3.06 \times 10^{-30} \ {\rm cm}^{-1} \, , \quad \, \lambda _0 \simeq 9.54 \times 10^{-54} \ {\rm cm}^{-2} \, .
\end{eqnarray}
The parameter $\lambda_0$ is best expressed as $\sim (100 {\rm Mpc})^{-2}$, suggesting that it is most important on scales of large galaxies or clusters. \\

We previously mentioned that the idea adopted to fit rotation curves resembles that of MOND, i.e. to introduce a new scale in the theory, which could be a scale intrinsically present in the data. Let us elaborate on this point more quantitatively. Considering the measured distance $R_{\text{last}}$ and rotational velocity $v_{\text{last}}$ of the outermost data in the rotation curves, it can be shown that the combination $\gamma_{\text{last}} \equiv v_{\text{last}}/c^2R_{\text{last}}$ for each of the galaxies in the sample is, within better than an order of magnitude, very close to the best-fit value for $\gamma$. In other words, the rotation curve data contain a preferred scale, which we introduced through the non-Newtonian correction. \\

We conclude this chapter by considering, for completeness, the case of a general power-law correction:
\begin{eqnarray}
\tilde a(r)=1-\frac{r_s}{r}+\gamma r^m-\lambda_0 r^2 \,,
\label{aex2}
\end{eqnarray}
with $m$ a positive real parameter, whit the full metric given by Eq.(\ref{eqb}), namely:
\begin{eqnarray}
\hspace{-1cm}b(r)=\frac{\left(4(2+m)r^2-12(2+m) r_s r+9(2+m)r_s^2-4(m-2)\gamma r^{2+m}\right)(r-r_s+\gamma r^{1+m}-\lambda_0 r^3)}{(2+m)r\left(3r_s-2r+(m-2)\gamma r^{1+m}\right)^2}\,.
\nonumber\\
\label{bm}
\end{eqnarray}
In deriving this last expression we have set the integration constant in such a way as to recover the Schwarzschild solution in the limit $\gamma=\lambda_0=0$, namely at short distances. On the other hand, at large distances, in the limit $r_s=\gamma=0$, we once more find the static patch of the de Sitter solution. Instead, at galactic scales, in the limit $r_s=\lambda_0=0$, we obtain:
\begin{eqnarray}
\tilde a(r)=1+\gamma^m r\,,\quad
b(r)\simeq
\frac{4(1+\gamma r^m)(2+m+2\gamma r^m-m\gamma r^m)}{\left(2+m)((m-2)\gamma r^m-2\right)^2} \ .
\end{eqnarray}
The potential can be found only in an implicit way and its on-shell expression results to be:
\begin{eqnarray}
V(r)=\frac{2m^2\gamma r^{m-2}\left(\gamma^2m^2r^{2m}+m(8-6\gamma r^m-4\gamma^2 r^{2m})+4(2+3\gamma r^m+\gamma^2 r^{2m})\right)}{(2+m)\left((m-2)\gamma r^m-2\right)^3} \ .
\label{vm}
\end{eqnarray}
The trivial case $m=2$ corresponds to a de Sitter-like solution with negative cosmological constant (for $0<\gamma$), :
\begin{eqnarray}
\tilde a(r)=b(r)=1+\gamma r^2 \,,\quad m=2\,.
\end{eqnarray}
In this case the field is given by $\phi= \pm i\text{arcsinh} [\sqrt{\gamma}r]/\sqrt{\gamma}$ and the potential reads $V(\phi)\simeq-6\gamma$.  \\

The non-Newtonian correction we have considered has recently been studied in the context of $f(R)$ gravity in \cite{capo}. In the low-energy limit of power-law $F(R) \propto R^n$ gravity, a $r^m$ correction to the Newtonian potential emerges, with $m$ related to the power $n$ as:
\begin{eqnarray}
m = \frac{12n^2-7n-1-\sqrt{36n^4+12n^3-83n^2+50n+1}}{6n^4-4n+2}
\end{eqnarray}
Stability of the potential at large distances and constraints from Solar System tests require $0 < m < 1$. In \cite{capo} the above model was fitted to 15 LSB galaxies, and an excellent fit is achieved, with best-fit value $m = 0.817$ (corresponding to $n = 3.5$). Interestingly, the study \cite{salucci} performs a similar fit to two objects in \cite{salucci} for which a particle dark matter explanation works fails to explain rotation curves successfully: the dwarf galaxy Orion and the low luminosity spiral NGC 3198 (which had not been analysed in \cite{capo}). In this case, an excellent fit is obtained for $m = 0.7$. \\

Let us conclude by making an important remark. Despite the potential leading to the chosen non-Newtonian corrections has only been given implicitly, it nonetheless the form of $b(r)$ has been provided explicitly, demonstrating how the desired behaviour for galaxy rotation curves, consistent with data, can be obtained in mimetic gravity. This is a notable result, as it implies that not only is it possible to reproduce the behaviour of dark matter on cosmological scales in mimetic gravity, but on astrophysical scales as well, solving the galaxy rotation curves problem, which is one of the pillars of evidence for dark matter.

\section{Conclusion}

Mimetic gravity has emerged as an interesting and viable alternative to General Relativity, wherein the dark components of the Universe (underlying dark matter, the late-time acceleration, and inflation) can find an unified geometrical explanation and interpretation. The theory is related to General Relativity by a singular disformal transformation, which is the reason behind its exhibiting a wider class of solutions. Here, we have reviewed the main aspects of mimetic gravity, beginning by placing it in the wider context of theories of modified gravity. After having reviewed the underlying theory behind mimetic gravity, we have studied some of its solutions and extensions, such as mimetic $f(R)$ gravity, mimetic unimodular gravity, and others, focusing on the reconstruction technique which allows the realization of numerous wishful cosmological expansion scenarios. After having discussed the issue of perturbations within the theory, we have considered a specific mimetic-like model, namely mimetic covariant Ho\v{r}ava-like gravity, wherein we applied the concepts discussed in the first part of the review. The final part of the review has been devoted to the study of static spherically symmetric solutions within mimetic gravity, and the application of these to the study of rotation curves within such theory. \\

The dark components of our Universe remain as mysterious as ever. It is possible that we might shed light on the nature of dark matter, dark energy, and inflation, as more data from experiments and surveys pours in the coming years. Thus far, theories of modified gravity, despite their \textit{prima facie} complications, appear as viable and theoretically well motivated routes to pursue. This is true for mimetic gravity as well, and thus we expect it to remain an active arena of research in the field of modified gravity in the coming years.

\section*{Acknowledgements}

S.V. thanks the Niels Bohr Institute, where the majority of this work was completed, for hospitality. We have benefited from many discussions with Sergei Odintsov and Sergio Zerbini, whom we wish to thank.

\section*{Disclosure policy}

The authors declare that there is no conflict of interest regarding the publication of this paper.

\end{document}